\documentclass{aa}
\usepackage{graphicx}
\usepackage{natbib}
\bibpunct{(}{)}{;}{a}{}{,} 
\usepackage{amsmath}
\usepackage[dvips,usenames]{color}

\newcommand{\abbrev}[1]{{#1}}      

\newcommand{\new}[1]{{#1}}

\newcommand{\refereeEVH}[1]{#1}
\newcommand{\referee}[1]{{#1}}

\newcommand{\ebv}{$E(B-V)$}
\newcommand{\dmod}{$(m-M)_0$}
\newcommand{\feh}{{\rm [Fe/H]}}
\newcommand{\mh}{{\rm [M/H]}}
\newcommand{\KK}{K}

\newcommand{\distfor}{{$20.74 \pm 0.11$}}
\newcommand{\distforkpc}{{$141$}}
\newcommand{\ebvfor}{{$0.03$}}

\newcommand{\shallow}{{\sc shallow}}
\newcommand{\deep}{{\sc deep}}

\newcommand{\realfigure}[3]{
              \begin{figure}
              \resizebox{\hsize}{!}{\includegraphics{#1}}
              \caption{#2}\label{#3}
              \end{figure}
              }

\newcommand{\mscred}{{\sc mscred}}
\newcommand{\wfpred}{{\sc wfpred}}
\newcommand{\daophot}{{\sc daophot\,ii}}
\newcommand{\allstar}{{\sc allstar}}

\newcommand{\caii}{\ion{Ca}{ii}}

\newcommand{\hi}{\ion{H}{i}}

\def\nodata{~~~--\ \null}

\begin{document}
%
   \title{Near-infrared observations of the Fornax dwarf 
        galaxy. I. The red giant branch
        \thanks{Based on data collected at the European Southern Observatory,
         La Silla, Chile, Proposals No. 65.N-0167, 66.B-0247.}
        }

   \author{M.~Gullieuszik\inst{1,2}
          \and
          E.~V.~Held\inst{1}
          \and
          L.~Rizzi\inst{3}
          \and
          I.~Saviane\inst{4}
          \and
          Y.~Momany\inst{1}
          \and
          S.~Ortolani\inst{2}
          }

   \offprints{marco.gullieuszik@oapd.inaf.it}

   \institute{
Osservatorio Astronomico di
Padova, INAF, vicolo dell'Osservatorio 5, I-35122 Padova, Italy
        \and
Dipartimento di Astronomia, Universit\`a di
Padova, vicolo dell'Osservatorio 2, I-35122 Padova, Italy
        \and
Institute for Astronomy, University of Hawaii,
2680 Woodlawn Drive, Honolulu, HI 96822, USA
        \and
European Southern Observatory, Casilla 19001,
Santiago 19, Chile
        }
   \date{Received \dots; accepted \dots}


\abstract
{}
{We present a study of the evolved stellar populations in the dwarf
spheroidal galaxy Fornax based on wide-area near-infrared observations,
aimed at obtaining new independent estimates of its distance and
metallicity distribution.  Assessing the reliability of near-infrared
methods is most important in view of future space- and ground-based deep
near-infrared imaging of resolved stellar systems.}
{We have obtained $JHK$ imaging photometry of the stellar
populations in  Fornax. The observations
cover an $18.5 \times 18.5$ arcmin$^2$ central area with a mosaic of
SOFI images at the ESO NTT. Our data
sample all the red giant branch (RGB) for the whole area.  Deeeper
observations reaching the red clump of helium-burning stars have also been
obtained for a $4.5 \times 4.5$ arcmin$^2$ region.}
{Near-infrared photometry led to measurements of the distance to Fornax
based on the $K$-band location of the RGB 
tip and the red clump. Once
corrected for the mean age of the stellar populations in the galaxy, the
derived distance modulus is \dmod\,$ = $\,\distfor, corresponding to a
distance of \distforkpc\ Kpc, in good agreement with estimates from
optical data.
\refereeEVH{ We have obtained a photometric estimate of the mean
metallicity of red giant stars in Fornax from their $(J-K)$ and $(V-K)$
colors, using several methods.}
The effect of the age-metallicity degeneracy on the combined
optical-infrared colors is shown to be less important than for optical
or infrared colors alone.  By taking age effects into account, we have
derived a distribution function of global metallicity \mh\ from
optical-infrared colors of individual stars.  Our photometric
Metallicity Distribution Function covers the range $-2.0 < \mh < -0.6$,
with a main peak at $\mh \simeq -0.9$ and a long tail of metal-poor
stars, and less metal-rich stars than derived by recent spectroscopy. If
metallicities from \caii\ triplet lines are correct, this result 
confirms a scenario of enhanced metal enrichment in the last
1-4 Gyr.}
{}


\maketitle

\section{Introduction}

Stellar populations in dwarf spheroidal galaxies (\abbrev{dSph}) are
important for our understanding of galaxy formation and evolution. Dwarf
spheroidals in the Local Group can be studied in detail, giving
strong constraints on the star formation history (\abbrev{SFH}) and
chemical evolution of these system.  
Galaxies of the dSph type all started forming stars at an old epoch
($>10$ Gyr), but in most cases this early stellar generation was later
followed by major star-formation episodes giving rise to significant or
even dominant intermediate age populations.
While old and intermediate age stellar populations of dSph galaxies have
been the subject of many studies in optical bands \citep[see, e.g., ][
and references therein]{greb2005,held2005}, they have been little
studied in the near-infrared (\abbrev{NIR}).

However, NIR bands have several advantages when studying evolved stars
in these stellar systems.
Infrared photometry of evolved low-mass and intermediate-mass stars on
the red giant branch (\abbrev{RGB}) and the helium-burning phases (e.g.,
red clump, \abbrev{RC}) can be used to derive the basic
properties of galaxies (distance, metallicity).  Techniques to measure
such properties from near-infrared observations are becoming
increasingly important since the NIR wavelength domain will be central
to future instrumentation (ELT adaptive optics, JWST, etc...).
There are advantages in using near-infrared photometry for RGB
stars, and even more so in combining optical and NIR data. As we will
show in this paper, the age-metallicity degeneracy affecting the color
of RGB stars (and therefore metallicity determinations) is much less
severe using optical-infrared colors.

\new{In order to explore the information contained in the near-infrared
spectral window, we have undertaken an imaging study of the evolved
stellar populations in Local Group dwarf galaxies. }
%
%
The first galaxy we consider is Fornax, 
one of the most interesting cases to study, being one of
the most massive and luminous dSph satellites of the Milky Way
\citep{mate1998}.  This galaxy was one of the first to provide evidence
of an intermediate age stellar population, probed by the presence of
luminous carbon star on the asymptotic giant branch (\abbrev{AGB})
\citep{aaromoul1980,aaromoul1985,azzo+1999}.  Another indicator of a
conspicuous intermediate-age stellar population is the well populated
red clump of helium-burning stars \citep{stet+1998,savi+2000}.  

A wide plume of main-sequence turnoff stars indicates that the star
formation history of Fornax has been continuous from galaxy's formation
up to recent times \citep{stet+1998,buon+1999,savi+2000,pont+2004}.
The luminosity of the brightest blue stars shows evidence that Fornax
has been forming stars at least up to 200 Myr ago, while it does not
contain any stars younger than 100 Myr \citep{stet+1998,savi+2000}.
An old population is also present, as shown by the
presence of an horizontal branch (HB) and RR-Lyrae
\citep{berswood2002,grec+2006mex}, 
although the blue part of the HB is poorly
populated, so that the old and metal-poor population must be small
\citep{stet+1998,buon+1999,savi+2000}.  
\refereeEVH{The picture emerging from the above
mentioned studies indicates that Fornax began forming stars } 
at the epoch of formation of the Galactic
globular clusters (\abbrev{GGCs}) $\sim 13$ Gyr ago. The star formation
rate was quite low in the first Gyr, and then increased rapidly; Fornax
formed most of its stars 4-10 Gyrs ago.
Given the presence of recent star formation, one would expect to
find some gas associated with the galaxy.
\cite{youn1999} searched for neutral hydrogen out to the tidal radius,
and found none.
More recent observations by \citet{bouc+2006} revealed an extended \hi\
cloud in the direction of Fornax that may (or may not) be within the
galaxy. 

Two studies of the chemical enrichment history of Fornax from the
spectra of red giant stars, using the \caii\ infrared lines equivalent
widths, have recently been presented \citep{pont+2004,batt+2006}.
\citet{pont+2004} found a metallicity distribution of Fornax centered at
$\feh = -0.9$ \citep[on the scale of ][ hereafter
\abbrev{CG97}]{carrgratt1997}, with a metal poor tail extending to $\feh
\simeq -2.0$ and a metal-rich population reaching $\feh \simeq -0.4$.
The derived age-metallicity relation is well described by a chemical
evolution model with a low effective yield, in which an initial rapid
enrichment is followed by a period of slower enrichment, reaching $\feh
\sim -1 $ about 3 Gyr ago. Then a high star formation rate produced an
acceleration, increasing the $\feh$ to a recent value of $\sim -0.4$
dex.  
\new{These results agree with those previously obtained for a few stars
by \citep{tols+2003} using high resolution spectroscopy.
In a large spectroscopic study of RGB stars in Fornax, \citet{batt+2006}
noted the lack of stars more metal poor than \feh$ = -2.7$ and confirmed
the presence of a metal-rich tail up to $\feh \sim -0.1$.  They also
found that the metal-rich stellar populations are more centrally
concentrated, having also a lower velocity dispersion than metal-poor
stars.

This paper presents the results of a near-infrared study of RGB and RC
stars in Fornax, aimed at obtaining new independent estimates of its
distance 
as well as information on its metallicity distribution function from a 
combination of optical and NIR photometry.
This study may represent a local example of \abbrev{NIR} studies of
more distant resolved stellar systems with future ground-based and space
instrumentation.  }
Section~\ref{s_obs} presents the observations and the reduction, with
special care for the mosaicing techniques, and provides photometric
catalogs.

Color-magnitude diagrams are presented in Sect.~\ref{s_cmd}. 
In Sect.~\ref{s_lf}, the luminosity functions of red giant stars and
helium-burning red clump stars are obtained and used to
provide new estimates of the distance to Fornax. The mean metallicity and
metallicity distribution of red giant stars is derived in
Sect.~\ref{s_met} from their $(V-K)$ color distribution 
and compared with recent spectroscopic work. 
Finally, our results are summarized and briefly discussed in
Sect.~\ref{s_sum}.

\section{Observations and data reduction}
\label{s_obs}

\subsection{Observations}

Near-infrared observations of Fornax were carried out in November
10--11, 2000, using the SOFI camera at the ESO NTT telescope.  The camera
employed a 1024$\times$1024 pixel Hawaii HgCdTe detector which was read in
Double Correlated mode. We used SOFI in Large Field mode, yielding a
pixel scale of 0\farcs29 pixel$^{-1}$ 
and a total field-of-view of
about $5\arcmin \times 5\arcmin$.

\begin{table} 
\caption{
Observation log.}
\label{t_log}
\begin{tabular}{l r@{ }l  c r r@{$\times$}l}
\hline\hline 
field &
\multicolumn{2}{c}{night}&
filter &
N &
DIT&
NDIT \\
\hline
\deep\  & 10 &Nov. 2000	& $J$   &  8       &  10&12\\
        & 10 &Nov. 2000	& $H$   &  8       &   5&24\\
        & 10 &Nov. 2000	& $\KK$ &  16      &   5&12\\
\shallow\ & 11 &Nov. 2000	& $J$   &  1       &   5&12\\
        & 11 &Nov. 2000	& $H$   &  1       &   2&30\\
        & 11 &Nov. 2000	& $\KK$ &  1       &   2&30\\
\hline
\end{tabular}
\end{table}

\begin{figure}
\includegraphics[width=8.5cm]{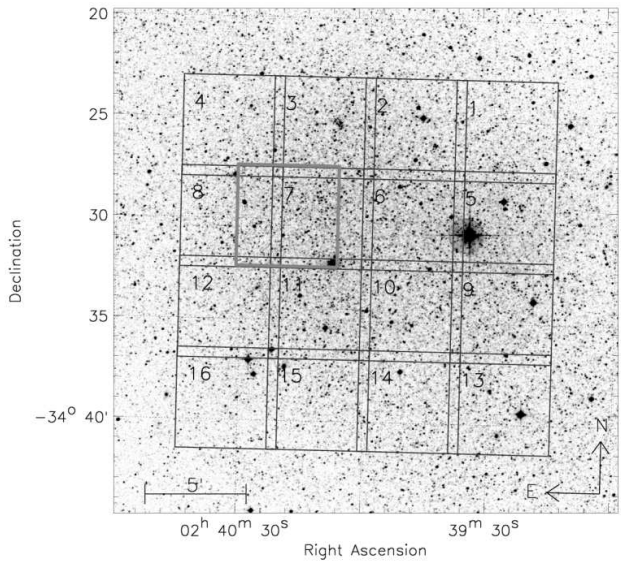}
\caption{
Map of the mosaic of 16 \shallow\ fields
observed in Fornax dSph, plotted on a  DSS2-red image. The location of the \deep\  field is shown as a {\it grey} box. }\label{f_map}
\end{figure}

Two complementary sets of images were taken in the $JHK_{\rm s}$
filters.  The first was a wide-area series of 16 \shallow\ contiguous
fields, in a square array of $4 \times 4$ covering about $18.5 \times
18.5$ arcmin$^2$. Secondly, we obtained a dithered sequence of images of
a central field of the galaxy providing \deep\ photometry for a $4.5
\times 4.5$ arcmin$^2$ area. We used an 8-points dithering pattern with
shifts of up to 10\arcsec\ from the central position.
The observing parameters are given in Table~\ref{t_log}, which lists the
number of images $N$ of each sequence along with the Detector
Integration Time (DIT) and the number of co-added integrations per image
(NDIT).  The on-target exposure times are 960s in $J,H,\KK$ for the
\deep\ image, and 60s for the \shallow\ mosaic.  The observation
strategy is further illustrated in Fig.~\ref{f_map}.

Given the non-negligible crowding of our Fornax field, we adopted an
observing strategy based on offset sky images alternated to 
on-target images. For the \deep\ image, the ``chopping'' time interval
varied between 120s for $J,H$ and 60 s for the $K$ band, 
while it was 60 s for
all bands for the \shallow\ observations. The offset pattern was a
dithered cross pattern for the \deep\ image and a simple offset in
declination for the \shallow\ scans.  Although expensive in terms of
observing time, this strategy proved to give extremely good sky
subtraction on scales comparable with the Point Spread Function
(\abbrev{PSF}) size, improving the quality of the photometry.

Standard stars (including very red stars) from \citet{pers+1998}
were observed on a regular basis during the nights at airmasses
comparable with those of target objects, to provide 
photometric calibration.  For each standard star, five images were
taken, with the star located at the center of the detector and in the
middle of the four quadrants of the frame.

\subsection{Reduction and astrometry}

The reduction steps were implemented in
IRAF\footnote{The Image Reduction and Analysis Facility (IRAF) software
is provided by the National Optical Astronomy Observatories (NOAO),
which is operated by the Association of Universities for Research in
Astronomy (AURA), Inc., under contract to the National Science
Foundation.}
as described in \citet{moma+2003}.
For the \shallow\ imaging, the basic observation and reduction unit was
one ``strip'' consisting of 4 \textit{science} and 4 \textit{sky}
frames.  The \textit{sky} images were scaled to a common median, after
rejecting the highest and lowest pixels, and averaged to produce a
master sky frame to be subtracted from all \textit{science} images of
the strip.  The reduction steps were similar for the \deep\ imaging,
using the four offset sky frames closest in time to each science image.
The sky-subtracted images were then corrected for bad pixels and
flat-fielded.  Illumination correction frames (as well as bad pixels
maps) available from the ESO Web pages were found suitable to this
purpose.
For the \deep\ imaging, all individual images, after sky subtraction,
correction of bad pixels, and flat-fielding, were registered and
averaged using the task {\sc imcombine} with rejection of the brightest
and faintest pixel for cleaning of cosmic ray hits.


\realfigure{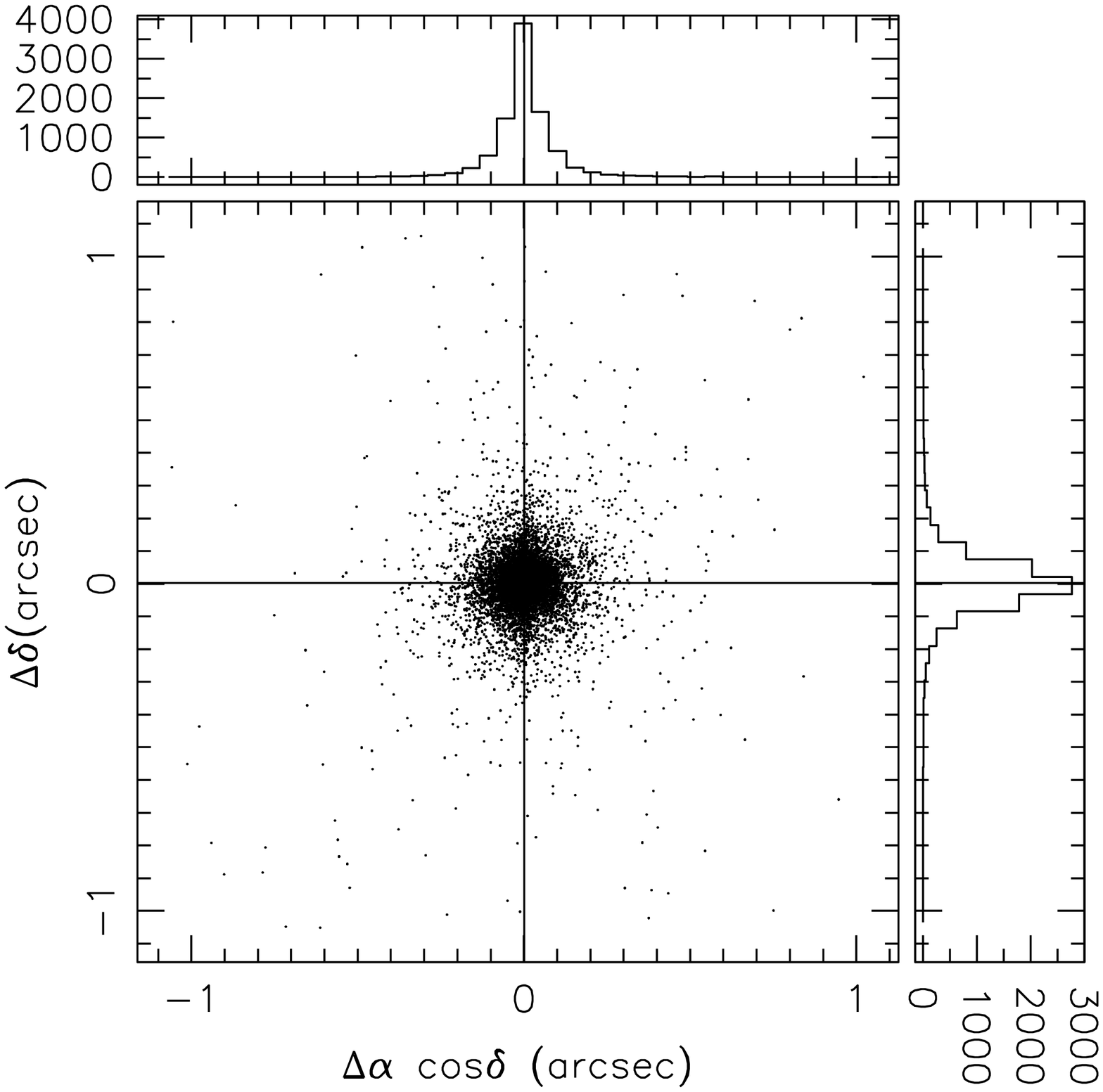} {Residuals in the ($\alpha, \,
\delta$) coordinates between the astrometric reference catalog (obtained
from ESO WFI observations) and the SOFI catalog, after astrometric
calibration. The data refer to the whole wide-area mosaic, composed of
16 fields.  The two histograms shows the $\Delta \alpha$, $\Delta
\delta$ marginal distributions. The mean shifts are negligible, 
$0\farcs002$ on both axes, while the standard deviation is $0\farcs10$ and
$0\farcs12$ on the right ascension and declination axis, respectively.
}{f_astrosol}

For the \shallow\ scans, mosaicing is a complex task because of the
modest overlap of the individual images.
We resorted to absolute astrometry to effectively
and accurately register the scans, match the photometry catalogs,
and produce a single mosaic image. To this purpose, we used the IRAF
package \mscred\ \citep{vald1998} along with the \wfpred\ script package
designed at the Padua Observatory by two of us (LR, EVH) for reduction
of  CCD mosaics.
The first step in the mosaicing was the characterization of the
astrometric properties of SOFI. This was done using a catalog of
secondary astrometric standards in the central region of Fornax
established from ESO-WFI observations \citep{rizz+2006}.  This secondary
catalog, in turn tied to the USNO-A2.0 reference system
\citep{monet+1998}, is required to provide a sufficiently high surface
density of stars needed to map distortions in the small field of SOFI.

The field distortion of SOFI was actually found to be negligible. 
Once a distortion map was constructed, all the individual exposures were
registered and resampled to a distortion-corrected coordinate grid with
a common reference World Coordinate System (WCS). As a result, a 
large-area mosaic image was reconstructed in each of the $JH\KK$ filters.

As a check of our astrometric calibration, we measured the right
ascension and declination of the secondary standard stars on the
WCS-calibrated $JH\KK$ frames. Figure~\ref{f_astrosol} shows the difference
between stellar coordinates on the SOFI wide-field catalog,
and the $\alpha, \delta$ of the same stars in the reference optical
catalog. 
This consistency test shows the {\it internal} precision of the
astrometric calibration, which is the really important figure for image
and catalog registration. The {\it absolute} (systematic) accuracy of
the coordinates is that of the reference catalog, estimated to be of the
order 0.2 arcsec.
The standard deviation of the residuals is not larger than 0\farcs12 on
both coordinates.

\subsection{Stellar photometry and calibration}
\label{sec_redu_photcalib}

Stellar photometry was obtained for the \deep\ and \shallow\ images
using \daophot/\allstar\ \citep{stet1987}. For the \deep\ photometry,
the \abbrev{PSFs} were generated from a list of isolated stars on the
coadded images. A {\sc Penny} function with a quadratic dependence on
position in the frame was adopted, to account for the elongation of the
stellar profiles especially affecting the left side of the chip (this
was traced to a misalignment of the Large Field objective). 
The photometry was finally performed on the stacked images using \allstar.
For the \shallow\ images, simple aperture photometry with an aperture
with radius equal to the FWHM of the stellar profile was found to yield
the best photometric precision.
For both catalogs, the pixel coordinates provided by \daophot\ were
converted to the equatorial system using the calculated WCS astrometric
calibration.

\realfigure{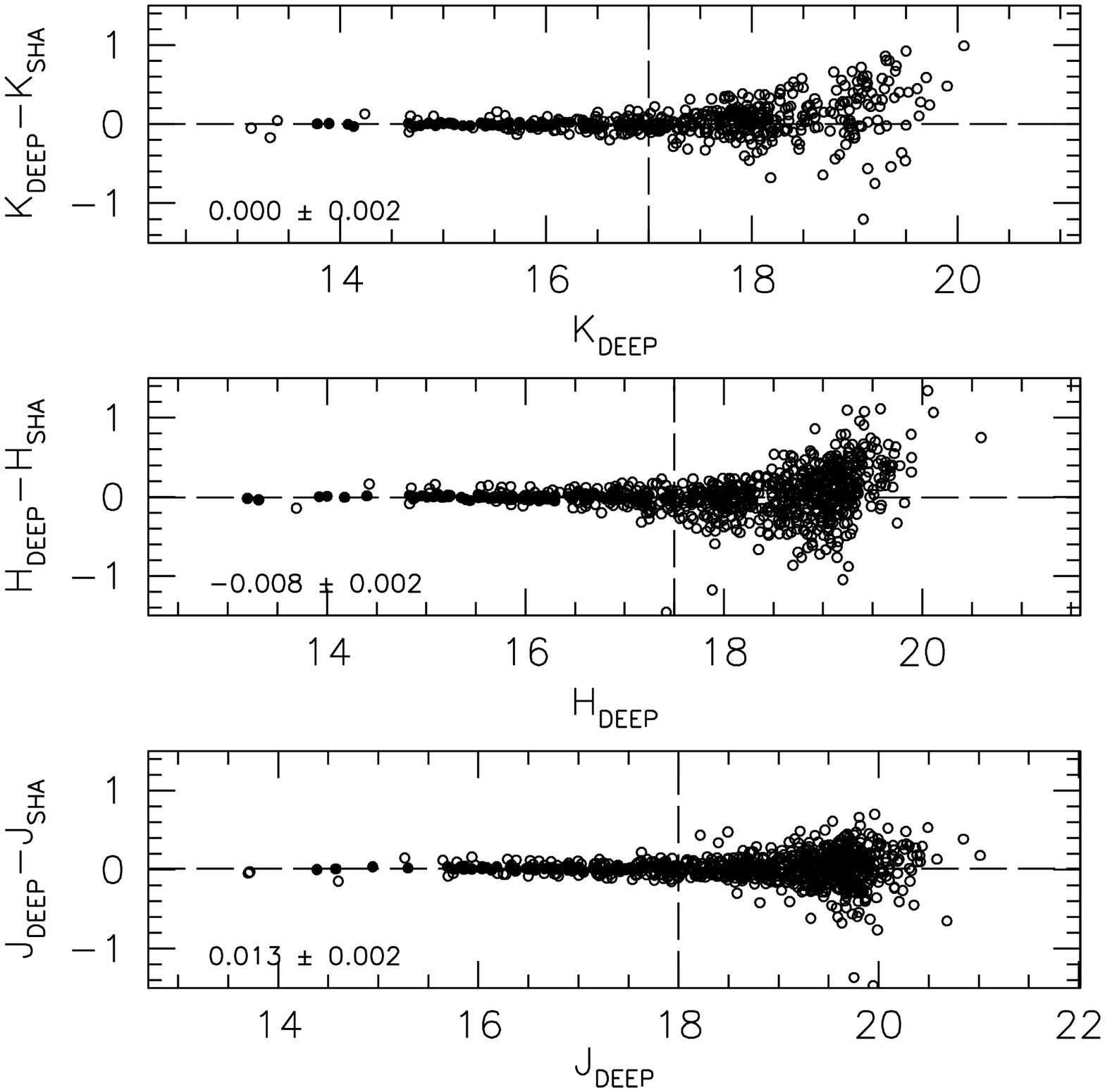}{Comparison of the \deep\ and
\shallow\ photometry of Fornax.}{f_depsha}

\realfigure{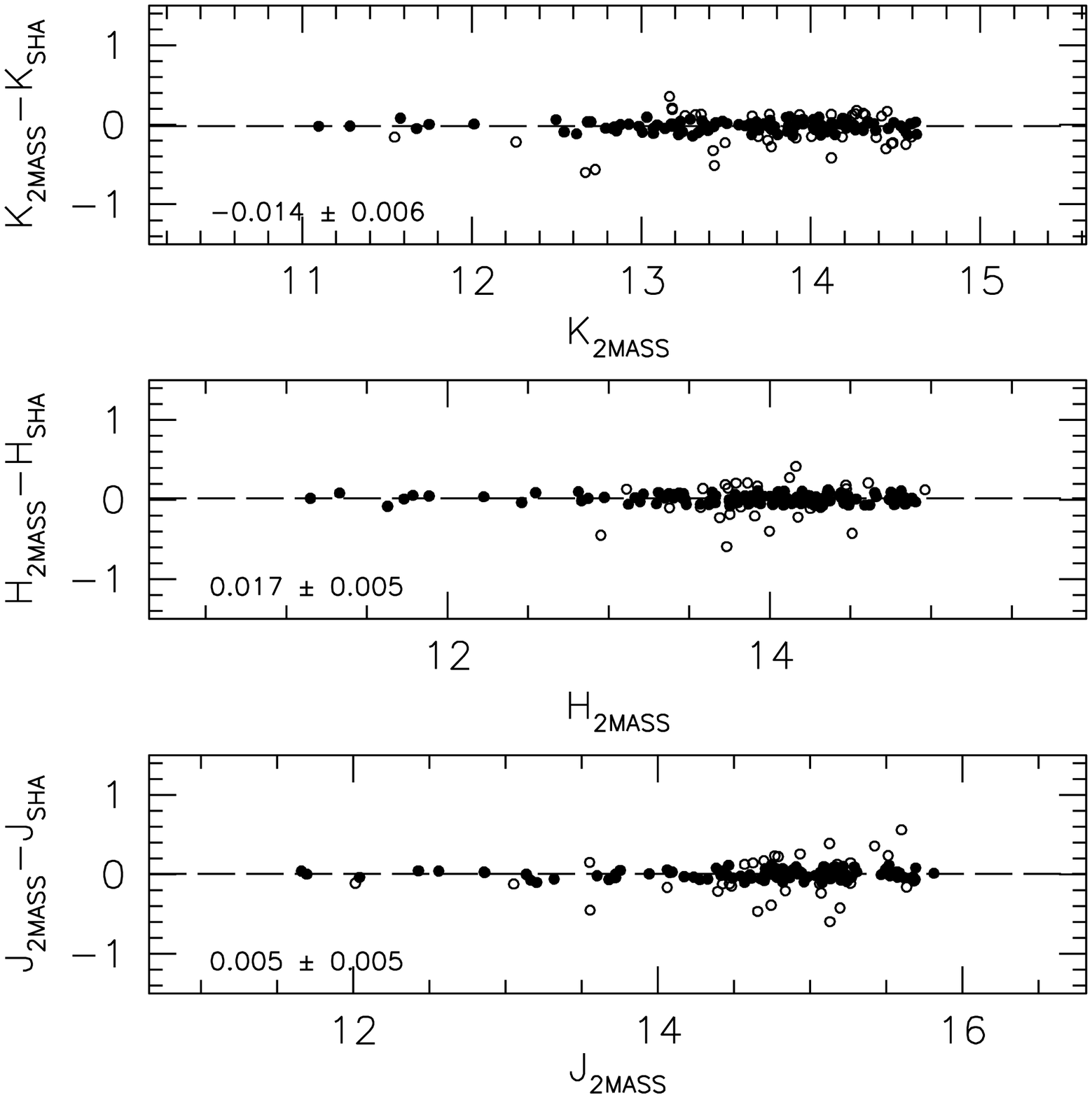}{Comparison of \shallow\ 
catalog with 2MASS photometry.  Filled circles are data that 
passed a $k\sigma$ rejection 
and were used to compute the mean differences.}{f_2msha}

The photometric calibration techniques are similar to those adopted by
\citet{moma+2003} and will be only briefly outlined here. 
Photometry of the standard stars through increasing apertures was
obtained with the IRAF {\sc apphot} task.  For the ``total'' magnitude,
we adopted a reference aperture of 18 pixel radius,  
close to the 10\arcsec\ aperture used by \citet{pers+1998}. The 5
measurements of each standard star were averaged after checking their
uniformity.
The instrumental magnitudes were finally normalized to 1~s exposure time
and zero airmass; for a magnitude $m_{\rm ap}$, calculated for an
observation with exposure time $t_{\rm exp}$ and airmass $X$, 
we defined the normalized magnitude $m$ as
\begin{equation}
m=m_{\rm ap}+2.5\, \log(t_{\rm exp})-\kappa_{\lambda }\, X \,
\end{equation}
where $\kappa_{\lambda }$ are the mean atmospheric extinction
coefficients:  $\kappa_{J}=0.08$, $\kappa_{H}=0.03$, and $\kappa_{K_{\rm
s}}=0.05$ were adopted from the ESO web page of SOFI. 

Given the small number of red stars observed each night, we used for
calibration all data available for five observing nights (including
data previously 
obtained in Feb. 2000 with the same SOFI setup).  All the
measurements were scaled to a common zero point 
and a first least square fit was done to compute the color terms of the
calibration relations.
Then, assuming fixed color terms, we measured the zero point variations
through the run (in particular between the two nights of Nov. 10, 11)
and found them to be very small, comparable with the measurements
errors; a single zero point was therefore adopted for the run, with
uncertainties 0.02 mag in $J$, $H$, and $K$. The
resulting calibration relations are:

\begin{equation}
\begin{split}
J-j \ \ =   -0.016\, (J-H) \; + \,23.118\\
H-h \   =   +0.001\, (J-H) \; + \,22.902\\
H-h \   =   -0.002\, (H-K)    + \,22.902\\
K - k_{\rm s}=+0.021\, (J-K) \; + \,22.337
\end{split}
\label{e_cal2}
\end{equation}
\noindent 
which are consistent with the calibration presented in the ESO web
page. 
\new{Note that our magnitudes on the \citet{pers+1998} $K$ system
are expected to show virtually no offset relative to the 2MASS system
\citep{carp2001}.  The color terms measured with respect to the
\citet{pers+1998} $JHK$ standard stars are negligible.}

The photometric catalogs of Fornax stars were calibrated using the
Eqs.~\eqref{e_cal2} after magnitude-scale and aperture correction.
This was based on {\sc apphot} large-aperture photometry and
growth-curve analysis of a few relatively bright, isolated stars in the
\deep\ and \shallow\ fields. The uncertainties on aperture correction
are of the order 
0.03 mag, yielding total calibration uncertainties 0.04 mag.

\refereeEVH{ In order to test the photometric calibration and rule out
the presence of photometric bias in the \shallow\ photometry, we compare
the zero points of the \deep\ and \shallow\ photometry in
Fig.~\ref{f_depsha}.  For the stars in common the photometry shows an
excellent internal consistency down to \KK~$\approx 18$, with zero point
differences of the order 0.01 mag.}
Figure~\ref{f_2msha} shows a comparison of our \shallow\ catalog with
2MASS photometry \citep{cutr+2003} on the same area. We have selected
only 2MASS measurements with $S/N>10$.  The systematic differences are
negligible considering the errors in the 2MASS catalog and the aperture
correction uncertainties.

\begin{table*}
\caption{The \shallow\ near-infrared catalog of Fornax stars over a
$18\farcm5 \times 18\farcm5$ area. A few lines are shown here for
guidance regarding its form and content, while the full catalog is
available from the CDS.}
\label{t_sha}
\begin{tabular}{lccccc}
\hline\hline       
ID & $\alpha$ & $\delta$ & $J$ & $H$ & $K$ \\
\hline                    
    235 & 2 40 08.59 &$-34$ 36 56.65     &  13.307 &  12.847 &  12.834 \\ 
   6117 & 2 39 54.21 &$-34$ 36 56.48     &  13.378 &  12.946 &  12.931 \\ 
   1114 & 2 40 02.16 &$-34$ 36 56.36     &  13.720 &  13.331 &  13.332 \\ 
   7263 & 2 40 22.48 &$-34$ 36 56.34     &  13.403 &  12.980 &  12.937 \\ 
   5036 & 2 40 01.84 &$-34$ 36 56.34     &  13.241 &  12.858 &  12.875 \\  
\hline                  
\end{tabular}
\end{table*}
%
\begin{table*}
\caption{The first rows of the \deep\ near-infrared catalog of stars in
a $4\farcm5 \times 4\farcm5$ central region of Fornax. The entire
catalog is available only electronically at the CDS.}
\label{t_deep}
\begin{tabular}{lccccc}
\hline\hline       
ID & $\alpha$ & $\delta$ & $J$ & $H$ & $K$  \\
\hline                    
    257 & 2 40 30.48 &$-34$ 31 26.27     &  13.703 &  13.199 &  13.135 \\ 
    818 & 2 40 30.45 &$-34$ 31 50.01     &  14.575 &  14.176 &  14.130 \\ 
   1544 & 2 40 30.40 &$-34$ 28 58.51     &  15.265 &  14.426 &  14.238 \\ 
    782 & 2 40 30.38 &$-34$ 31 43.16     &  16.251 &  15.706 &  15.634 \\ 
   1728 & 2 40 30.35 &$-34$ 28 32.03     &  15.936 &  15.178 &  15.066 \\ 
\hline                  
\end{tabular}
\end{table*}

The photometry lists obtained for the individual pointings were merged
into a single \shallow\ catalog. The stars were matched on the basis of
right ascension and declination, using an error box of 1\farcs5.
Finally, the coordinates of stars in the two catalogs were matched to
those of $BVI$ photometry (selecting objects with $V<23$) from ESO WFI
observations \citep{rizz+2006}, thus producing a final list with $BVIJHK$
magnitudes.
The \shallow\ and \deep\ catalogs of near-infrared photometry of Fornax
stars are presented in Tables~\ref{t_sha} and \ref{t_deep} (the entire
catalogs are made available electronically at the CDS).

\subsection{Artificial star experiments}\label{s_compl}

\realfigure{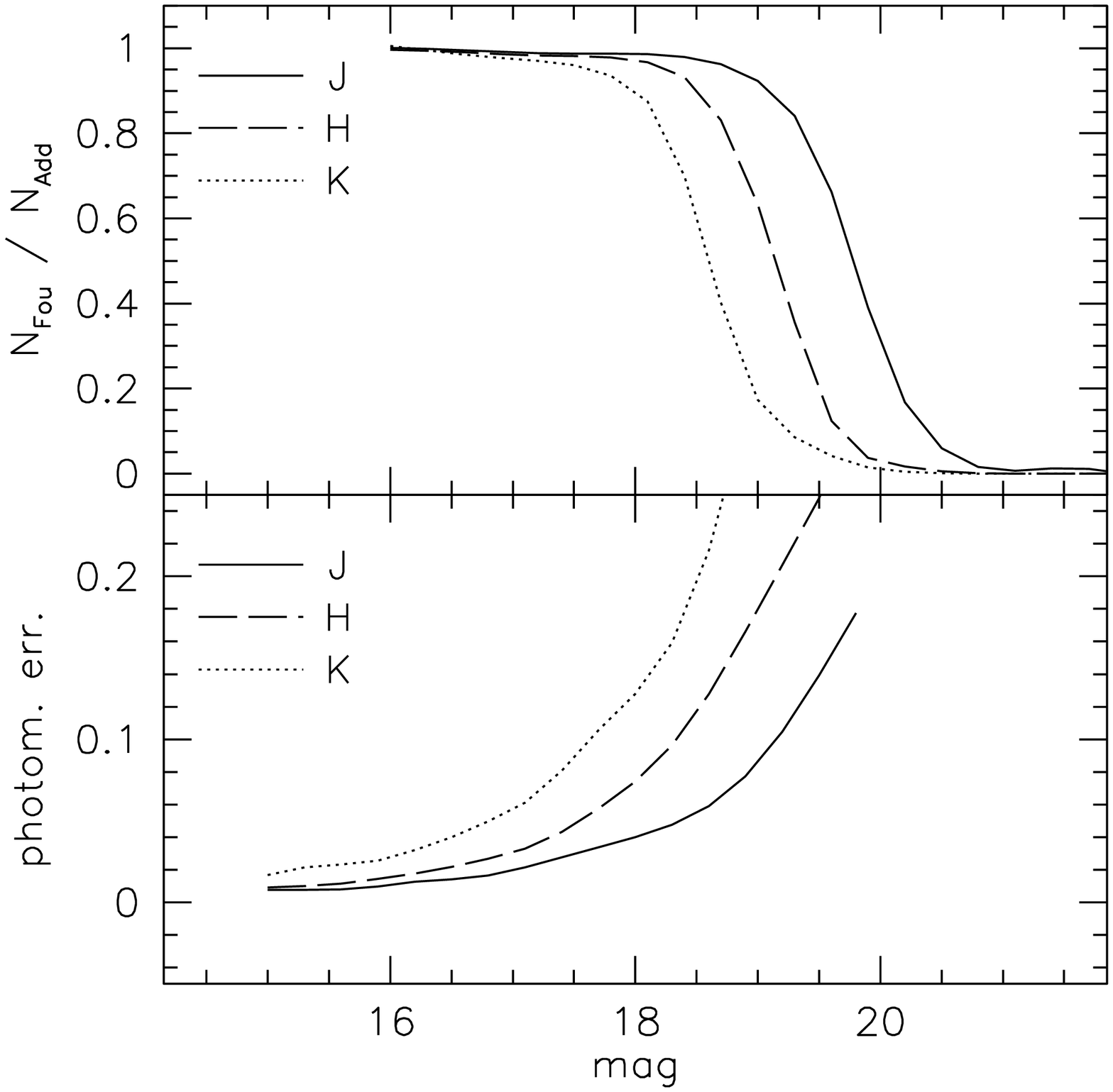}{
Completeness and photometric errors from 
artificial star experiments for the \shallow\ photometry.
{\it Upper panel:} 
Completeness calculated as the number of artificial stars recovered
as a function of observed magnitude.
{\it Lower panel:} 
photometric errors, i.e. the standard deviation of 
the differences between input and output magnitudes.
}{f_compl_s}
\realfigure{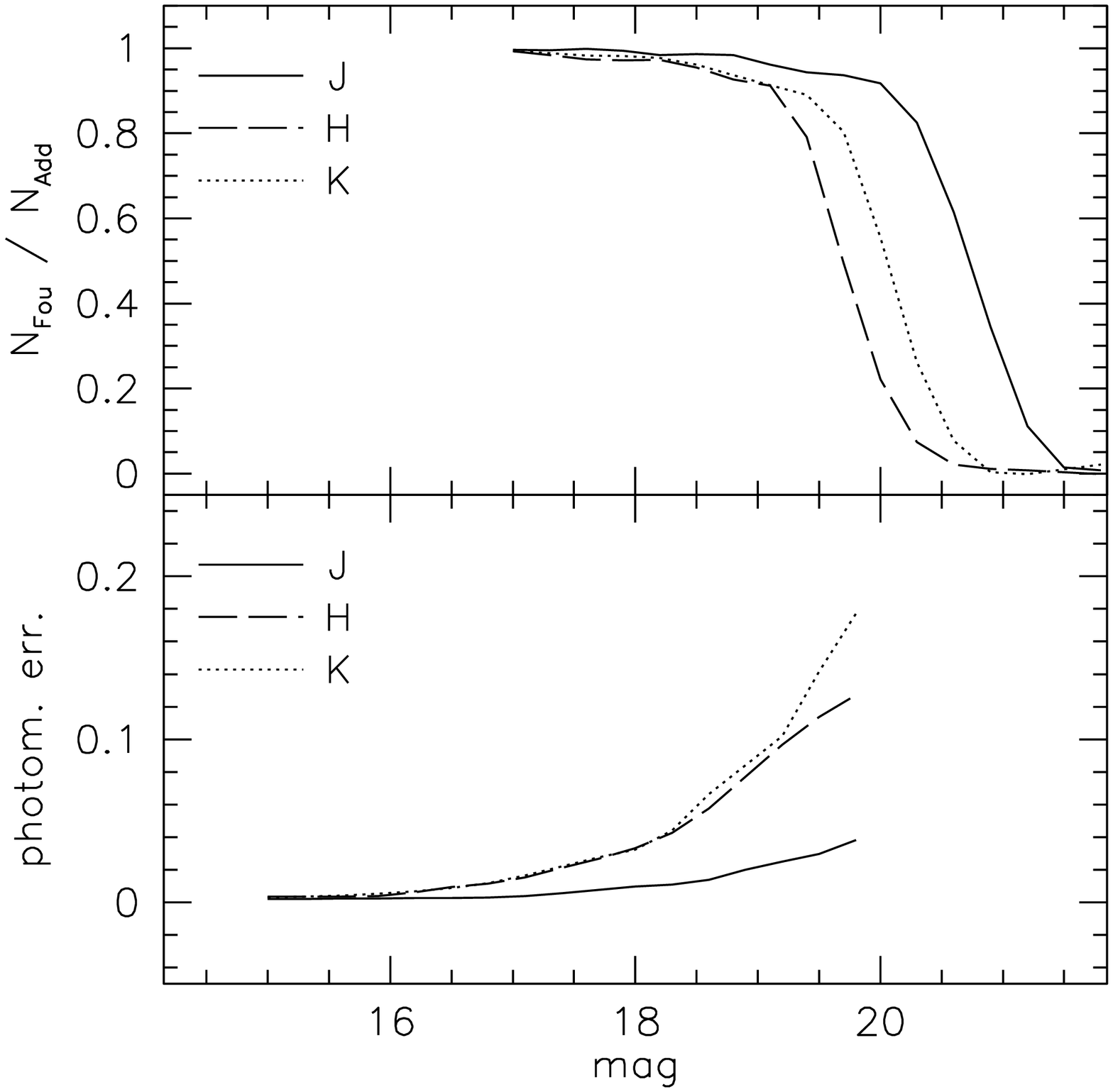}{The results of artificial star
experiments for the \deep\ photometry.}{f_compl_d}

\referee{
The completeness of our photometric catalogues was evaluated from
artificial star experiments.  We performed 
20 test runs by adding 400 stars to
the scientific frames in each run.
Since the crowding is uniformly low over the 16 \shallow\ pointings, with
only about 1000 stars per pointing on average, the experiments were
limited to one of the \shallow\ pointings.
%
The input magnitudes and colors of the artificial stars were chosen
along a sequence corresponding to the Fornax RGB.
The stars were
placed on a grid of equilateral triangles with a side of 40 pixels
($\sim 11\farcs5$), much larger than the stellar PSF.  In each
experiment the grid was then randomly shifted in order to uniformly
cover all the frame.
The results from artificial star experiments on the \shallow\ and
\deep\ photometry are shown in Fig.~\ref{f_compl_s} and
Fig.~\ref{f_compl_d}, respectively. We note that the \deep\ data are
comparatively rather noisy in the $H$ band with respect to the $J$ and
\KK\ bands.}

\referee{Another well-known observational effect is the fact that the
mean difference between input and output magnitudes is generally biased
towards brighter output magnitudes \citep[e.g.,][]{gall+1996}.  In our
experiments for the \shallow\ data, the mean difference of input and
output magnitudes resulted to be less than 0.05 mag 
for $K < 18.2$.  This is also an upper limit for the $(J-K)$ color
shift down to this limiting magnitude.  However, the color bias is
negligible ($< 0.01$ mag) in the first 2 mag below the RGB tip used for
the metallicity measurements.  }


\section{Color-magnitude diagrams}
\label{s_cmd}

\begin{figure*}[ht]
\centering
\includegraphics[width=8.cm]{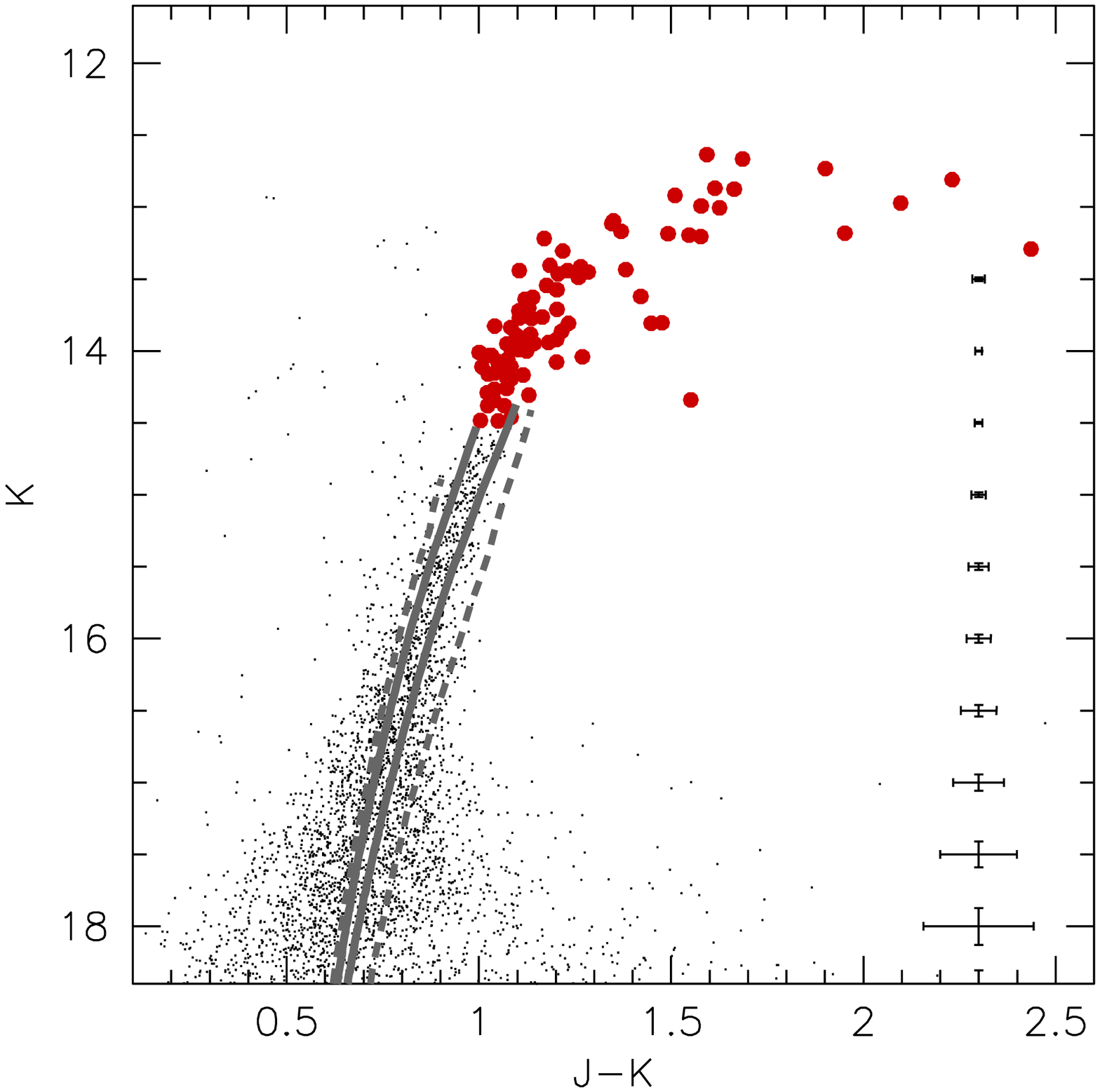}
\includegraphics[width=8.cm]{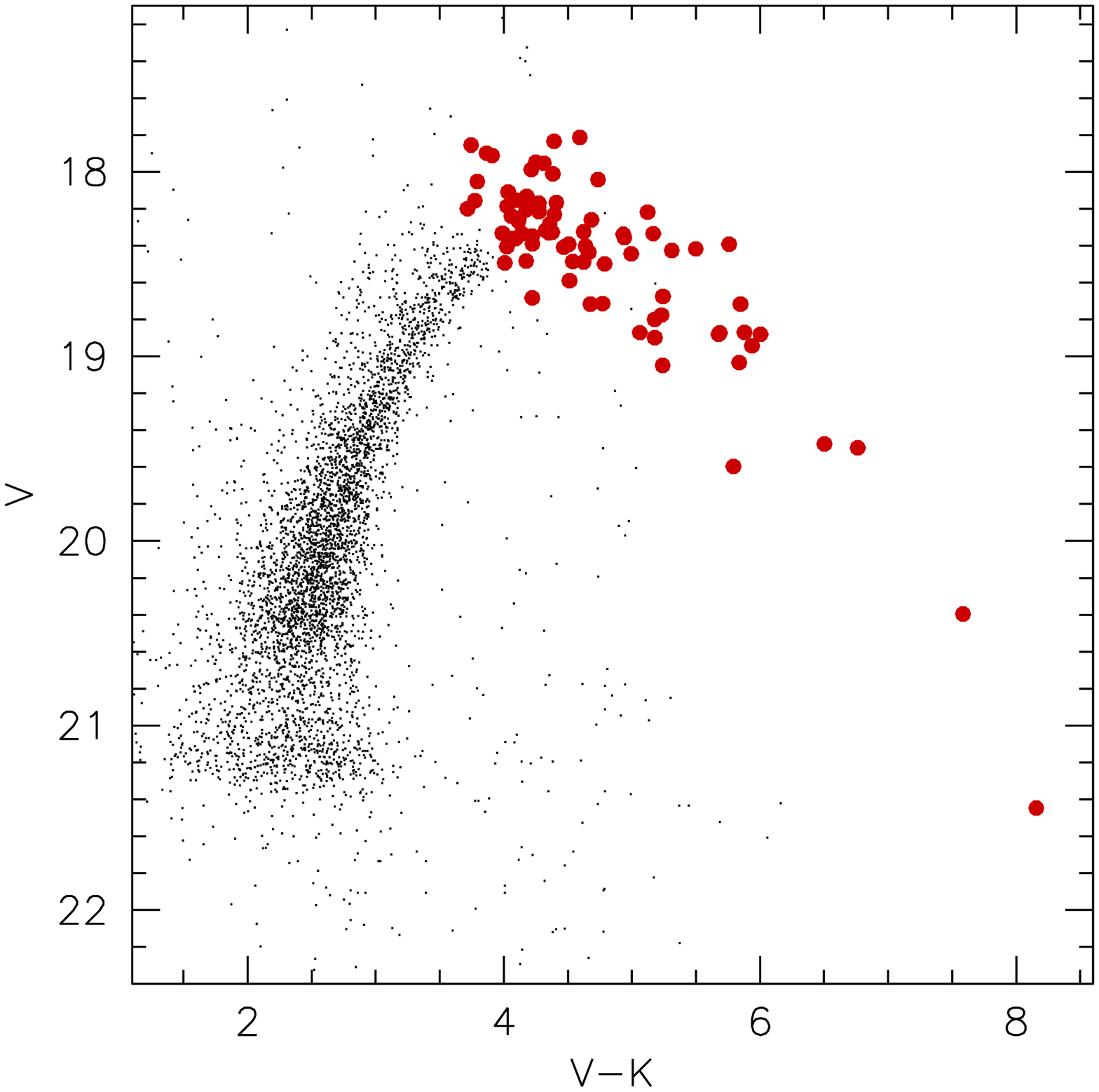}
\includegraphics[width=8.cm]{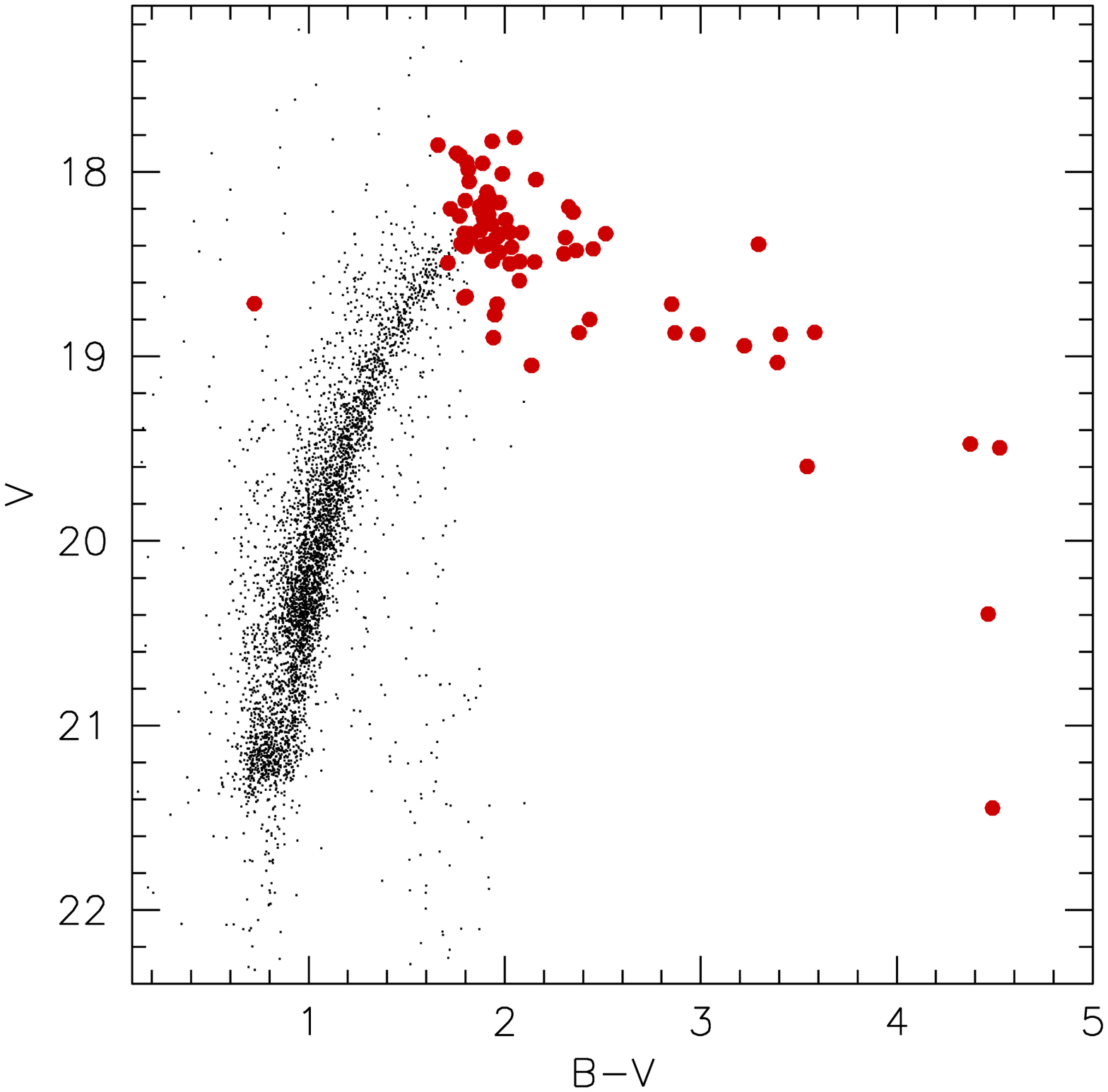}
\includegraphics[width=8.cm]{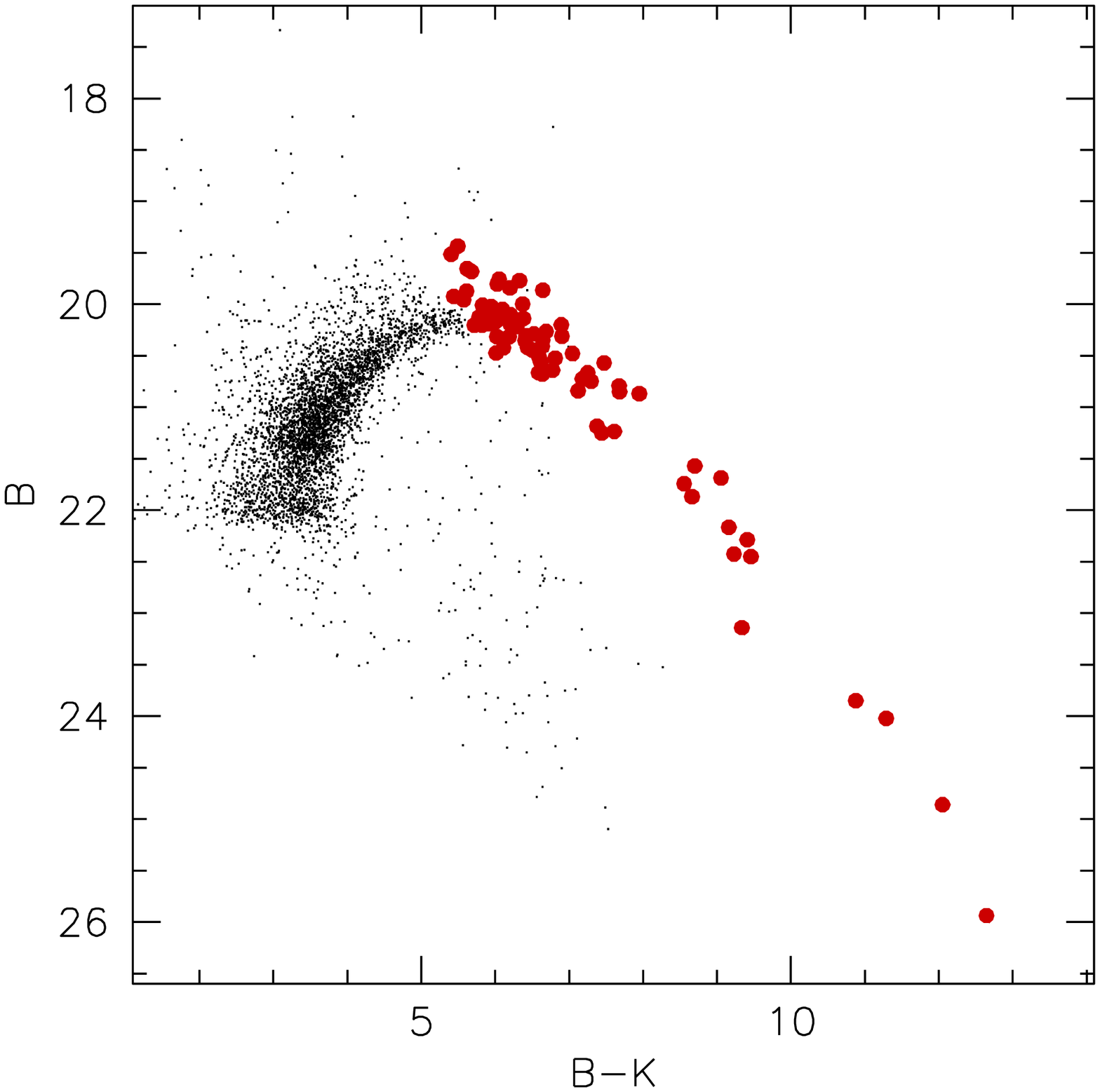}
\caption{
The color-magnitude diagrams of Fornax from 
the \shallow\ catalog,
obtained by combining optical and near-infrared photometry.  Stars on
the extended AGB, chosen redder than $(J-K) = 1$ and brighter than $K =
14.5$, are shown as {\it filled circles}.  Superimposed on the $K$,
$(J-K)$ diagram are theoretical isochrones from
\citet[\null:~G02, {\it dashed lines}]{gira+2002} and
\citet[\null:~P04, {\it solid lines}]{piet+2004}. The isochrone age
is 7 Gyr and the metal abundances are the closest to the metallicity of
Fornax for each model set: $Z=0.001$ and $0.004$ for G02, $Z=0.001$ and
$0.002$ for P04.  Typical photometric errors \referee{as derived from artificial star experiments} are also shown as crosses in the upper-left panel.}
\label{f_bvijhk}
\end{figure*}

The near-infrared and optical-infrared color-magnitude diagrams 
(\abbrev{CMDs}) of Fornax
\abbrev{dSph} in the \shallow\ sample are presented in
Fig.~\ref{f_bvijhk}.  
\referee{ Given the better image quality and spatial resolution of our
optical photometry, we used the \allstar\ shape parameter {\it sharp}
for the $V$ measurements to remove bad and non-stellar objects from our
photometric catalogues. Only objects with $-0.3 < $ {\it sharp} $< 0.3$
were selected.
}

The most noteworthy feature in the CMD in Fig.~\ref{f_bvijhk} is the
well defined sequence of intermediate-age AGB stars.
A fraction of AGB stars have red colors typical of carbon (\abbrev{C})
stars, and a few stars show extremely red colors possibly indicating
dust-enshrouded AGB stars. The properties of near-infrared selected AGB
stars in Fornax will be discussed in more detail in a separate paper.
We only remark here the dramatic change in their distribution when
optical and near-infrared bands are used. Note, for example, that
the redder AGB stars become progressively fainter in the bluer optical
bands, so that even the brightest (in terms of bolometric luminosity) 
are missed in optical CMDs. For the reddest
AGB stars the $(B-V)$ colors saturate.
Thus, selection in the near-infrared appears to be extremely important
to investigate the evolution of luminous AGB stars.  Also, an advantage
of optical--near-infrared colors is the improved discrimination against 
field contamination by foreground stars and background galaxies, 
as is evident, for 
example, when comparing the $V$, $(B-V)$ and $K$, $(J-K)$ diagrams in
Fig.~\ref{f_bvijhk}. 

The shallow, large-area catalog provides the statistics for studying the
AGB stars, the RGB tip, and the stellar populations of Fornax.  The
fainter limiting magnitude and better precision of our \deep\ photometry
of a smaller central region allows us to characterize the RGB down to
the so-called ``AGB bump'' and the red clump.
Our \deep\ color-magnitude diagram is presented in
Fig.~\ref{f_IRdeep}, together with the RGB fiducial lines of Milky Way
globular clusters \citep{vale+2004a}. This diagram is similar to that
obtained by \citet{pietrz+2003} with ISAAC at the VLT. 
Note that the RGB of Fornax is relatively thin in this small central
field.  It lies between the fiducial RGB lines of M\,4
([Fe/H]~$= -1.19$ on the \citealt{carrgratt1997} abundance scale) and
M\,107 ([Fe/H]~$= -0.87$). This suggests a metallicity [Fe/H]~$\sim -1$
on the same scale.

\realfigure{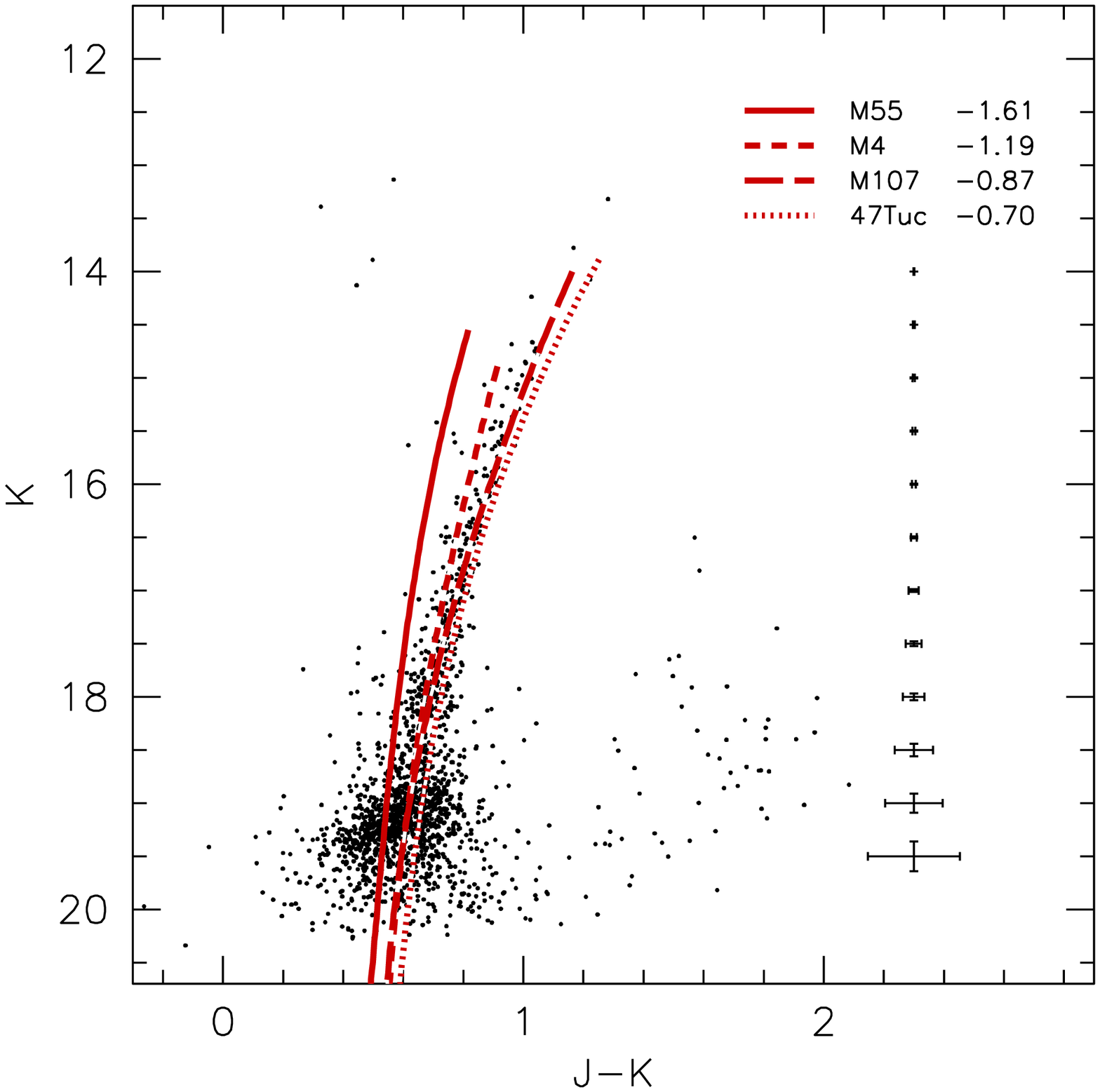}{
The near-infrared CMD of Fornax derived
from our \deep\ photometry, showing the RGB and the prominent ``red
clump'' of intermediate-age He-burning stars. Superposed on the RGB are
the fiducial lines of Galactic globular clusters from
\citet{vale+2004a}, with metallicities on the scale of
\citet{carrgratt1997}. 
Typical photometric errors \referee{as derived from artificial star experiments}
 are shown as crosses.
}{f_IRdeep}

\section{Luminosity function and distance}
\label{s_lf}

\realfigure{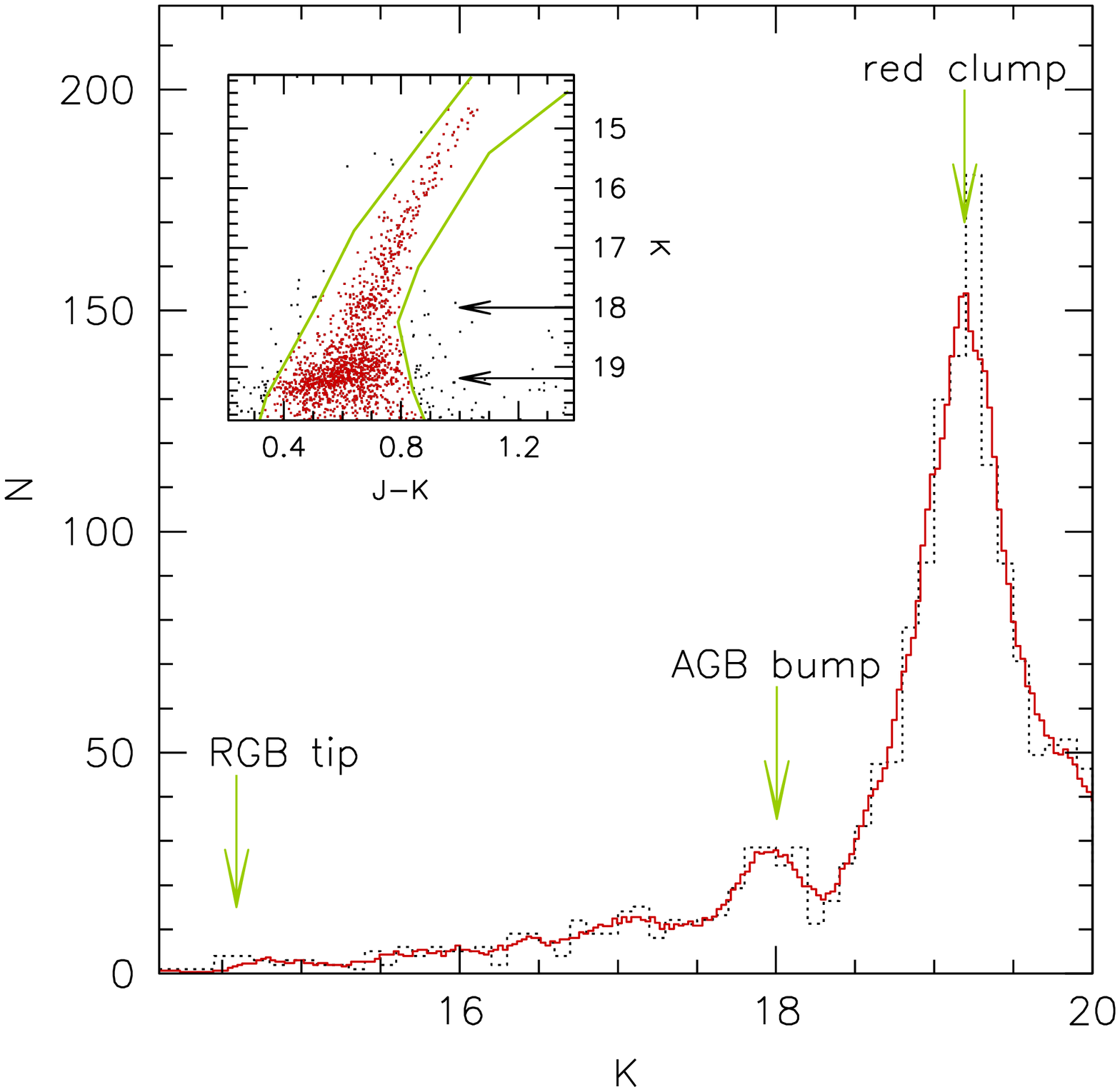}{The $K$-band luminosity function of
evolved stars in Fornax,
\referee{corrected for completeness effects.}
This LF, derived from the \deep\ catalog, shows
the location of the red clump and the AGB bump. The inset illustrates
our sample selection. A histogram with a 0.1 mag bin is shown as a {\it
dotted line}, while the {\it solid line} represents a multibin
histogram.
}{f_agb_rc}

Using the \deep\ data shown in Fig.~\ref{f_IRdeep}, we constructed a
luminosity function (\abbrev{LF}) extended down to the red clump of
Fornax (Fig.~\ref{f_agb_rc}).  \referee{ The LF was corrected by taking
into account the incompleteness factor derived in Sect. \ref{s_compl}.
}
Our sample of RGB and He-burning stars was
selected as shown in the inset. Using this LF, new estimates of the
distance to Fornax have been obtained based on the mean $K$-band
luminosity of the red clump and the $K$ magnitude of the RGB tip.

\subsection{The red clump}

Several authors have pointed out that distance measurements based on the
magnitude of the RC are most reliable in the near-infrared, because of
the smaller dependence of the RC mean luminosity on metallicity and age
in the $K$-band, along with the reduced reddening relative to the
optical wavelengths \citep[e.g., ][]{alve2000,groc+sara2002,pietrz+2003}.
%
\citet{pietrz+2003} employed VLT-ISAAC photometry of RC stars to measure
the Fornax distance. An independent distance estimate based on our own
measurement of $K_{\rm RC}$ is presented here, and the two
determinations compared.

In order to reduce the dependence from the bin choice, the LF was
constructed as a multibin histogram by averaging 10 LFs with a fixed 0.1
mag bin and starting points shifted in steps of 
0.01 mag.
A mean level 
$K_{\rm RC}=19.20 \pm 0.02$ 
was measured following the
standard procedure, i.e. by fitting the sum of a Gaussian and a
polynomial to the magnitude distribution of stars in the color range
$0.30 < J-K < 0.85$ \citep[see, e.g.,][]{pietrz+2003}. 
The mean
$K_{\rm RC}$ value and its standard deviation are the results of 5000
experiments with bootstrap resampling of the luminosity function. 
\referee{Based on our artificial star experiments, any magnitude shift
due to a photometric bias is negligible ($\sim 0.01$ mag) for the \deep\
data at the RC level.
Other sources of error are considered in the following, in
addition to the 0.02 mag fitting error, to evaluate the total
uncertainty in the distance to Fornax dSph.}

\new{ 
In order to compute the distance, our magnitudes \citep[tied to the LCO
system of ][]{pers+1998} need to be transformed onto the photometric
system used in the RC luminosity calibration of \citet{alve2000}.
Since our $K$-band photometry agrees very well with the 2MASS system
\citep[Sect.~\ref{sec_redu_photcalib}; see also][]{carp2001}, we simply
adopted the transformation from 2MASS to the \citet{bessbret1988}
system, $K_{\rm BB} = K_{\rm 2MASS}+0.044$.  The latter is very close to
the \citet{koor1983} system used by \citet{alve2000}
\citep[see][]{carp2001,groc+sara2002}.
}

We therefore applied the relation
\begin{equation}
  (m-M)_0 = (K_{\rm RC} + 0.044) - M_K - A_K + \Delta M_K
\end{equation}
\noindent 
where the $K$-band luminosity of the red clump in the solar neighborhood
\new{is $M_K=-1.61$ according to 
the \citet{alve2000} calibration (a value confirmed by
\citealt{groc+sara2002} from a sample of 14 open clusters), }
and $\Delta M_K$ is a population correction term.
This correction accounts for the
different stellar content of Fornax and the local Galactic RC on
which the \citet{alve2000} calibration is based.
The population correction was calculated using the precepts of
\cite{sala+gira2002}, the age-metallicity relation of \cite{pont+2004},
and the star-formation history by \cite{tols+2003}, and found to be
$\Delta M_K=-0.10$ (the RC in Fornax being fainter).
Adopting the extinction law of \citet{riek+lebo1985} and 
\ebv~$=$~\ebvfor,
the resulting distance is \dmod~$=20.84+\Delta M_K=$\,\distfor, 
where the uncertainty includes the statistical 
error on the RC location, the photometric zero-point error, 
\referee{and a photometric error of 0.1 mag at the level of the RC 
(see Fig. \ref{f_compl_d}).}
\citet{pietrz+2003}, considering the population correction negligible, 
found a distance modulus \dmod~$= 20.86$, which differs from our value
only for the correction term, the $K_{\rm RC}$ determination being in
perfect agreement.

\subsection{The AGB bump}

In the LF presented in Fig.~\ref{f_agb_rc} another bump is clearly seen.
Its $\KK$ band magnitude was measured by fitting also in this case the
sum of a polynomial and a Gaussian function and found to be $\KK = 17.99
\pm 0.03$
with a bootstrap technique similar to that used for the red clump.  This
feature is identified with the AGB bump, which is the signature in the
CMD of the beginning of the AGB phase. At the beginning of this
evolutionary phase the increase of luminosity of a star is slower than
in subsequent AGB phase, and thus a bump in the
\new{LF is produced \citep{v.cast+1991,gall1998,alvesara1999}.}

We also explored the possible identification of this feature with the
RGB bump.  using the \cite{vale+2004b} calibration of the RGB bump.
Assuming a metallicity \mh$=-1.6$ for the old population of Fornax
\citep{berswood2002,grec+2006mex}, we derived that the expected RGB bump
magnitude is $\KK=18.52$. As discussed by \cite{savi+2000}, the RGB bump
is fainter that the bump in Fig. \ref{f_agb_rc} and not visible because
it is too close to the overwhelming red clump.  We note that the
metallicity assumed here, \mh$=-1.6$, is appropriate for the old
population of Fornax and a lower limit to the actual mean metallicity of
intermediate age stellar populations (see following sections).
The RGB bump magnitude derived by \cite{vale+2004b} becomes fainter at
increasing metallicity, and therefore our estimate of the RGB bump
magnitude is to be considered a lower limit.  We therefore identify the
feature observed at $\KK = 17.99$ with the AGB bump.

\subsection{The RGB tip}

\realfigure{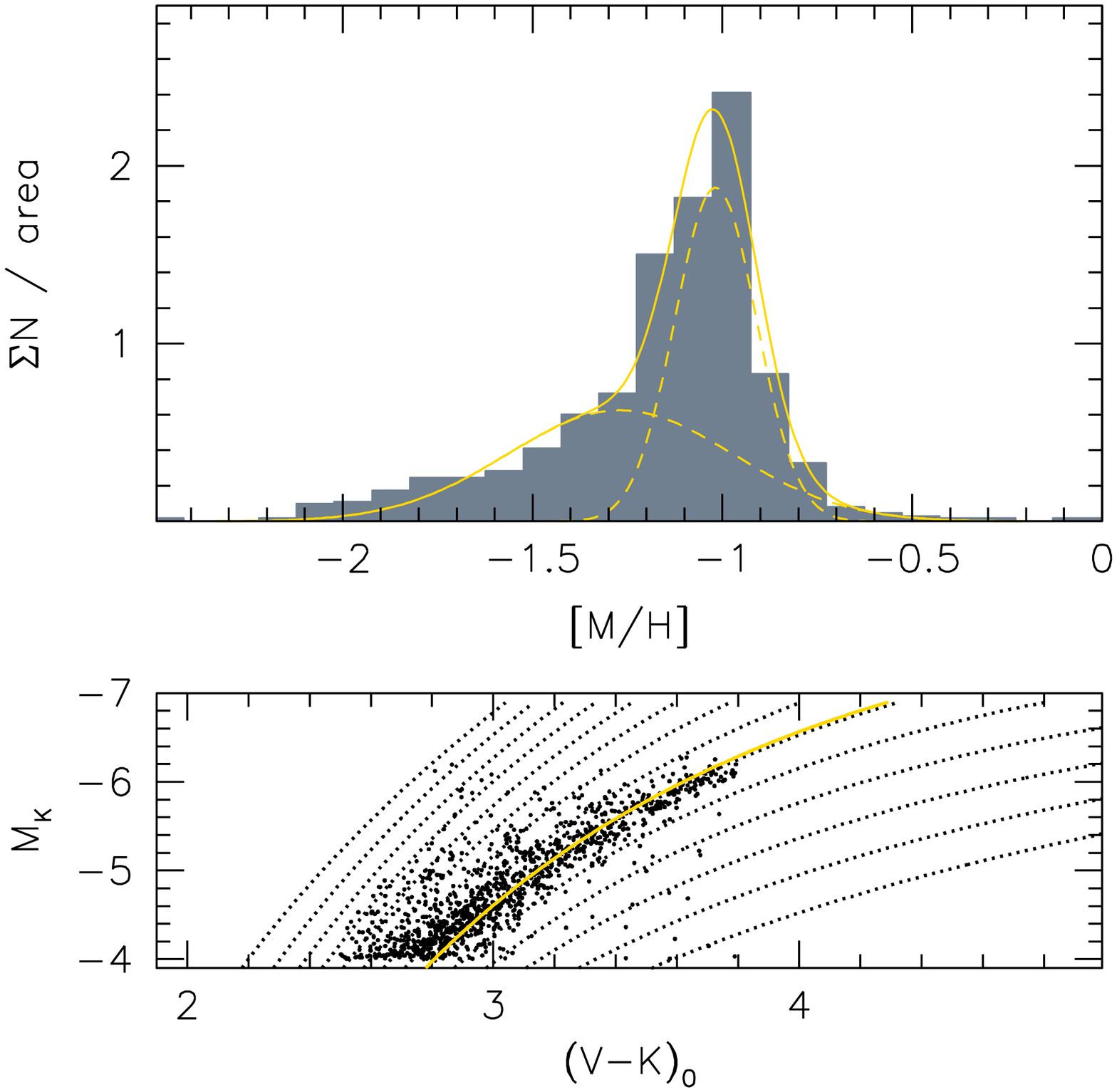}{The luminosity function of the upper RGB,
from the \shallow\ catalog.
The inset shows how RGB stars have been
selected.  The RGB cutoff was measured by fitting an error-convolved
step function ({\it green long-dashed line}) to a multibin histogram
({\it continuous line}). The {\it dotted line} shows an example of a
simple LF with 0.1 mag bins.  }{f_trgb}

Figure~\ref{f_trgb} shows a close-up view of the brighter part
of the RGB luminosity function in Fornax, obtained by selecting 
red giant stars from the wide field \shallow\ catalog
This catalog  was chosen
because of the better statistics, and, indeed, the RGB cutoff appears
very well defined.  We obtained an objective estimate of the magnitude
of the RGB tip by fitting the \abbrev{LF} with the
convolution of a step function with a Gaussian kernel representative of 
the measurement errors, as in \citet{moma+2002}. 
The function is composed of a constant value 
\new{brighter than the RGB cutoff, and a power law (on a log$N$ scale)}
below the RGB cutoff.
\referee{ 
Since the tip is located at \KK~$\simeq 14.6$ where our photometry is
complete (see Fig. \ref{f_compl_s}), no completeness
correction was applied to the LF.}
As before, a multibin LF was used, obtained by averaging 10 LFs
with a fixed 0.1 mag bin and intervals shifted by 0.01 mag.
Using this procedure the RGB tip was detected at $K=14.59 \pm 0.03$.
The uncertainty includes the fitting error and the error associated
with the binning of the LF. The internal measurement error at the level
of the RGB tip gives a minor contribution ($\sim 0.01$ mag).

An independent measurement of the RGB tip was also 
obtained using the Maximum
Likelihood Algorithm described by \citet{maka+2006}. The TRGB was found
at $K=14.63 \pm 0.03$.  Since the two independent measurements agree
within the errors, we adopt the mean value 
$K=14.61 \pm 0.02$~(random)~$\pm 0.03$~(systematic) 
as our final measure of the tip, where the systematic uncertainty 
reflects the photometric zero point  error (Sect.~\ref{s_obs}).

The age and metallicity dependence of the RGB tip is larger in $K$
than in the $I$-band, so that the application of the RGB tip method is
more uncertain in this case \citep[e.g., ][]{salagira2005}.
Intermediate-age stars show a fainter RGB tip than old stars, with a
difference that can be as large as $\sim 0.1$ mag in $K$ in the age
interval 4--13 Gyr.  On the other hand, if younger populations become
more metal-rich as a result of galactic chemical evolution, their RGB
tip becomes brighter, because the TRGB $K$ luminosity rises with
increasing metallicity.

Since the bulk of stellar populations in Fornax is of intermediate age,
a population correction is required to compute the distance. 
To this aim, we modeled the combined effects of age and metallicity on the
RGB tip by constructing a synthetic CMD containing 100\,000 stars in the
upper RGB. Our simulations are based on the ZVAR code 
\citep{bert+1992}, the Padova isochrones
\citep{gira+2002}, and adopt the chemical evolution history by
\cite{pont+2004} and the SFH from \cite{tols+2003}.
We obtained the population correction for the luminosity of the RGB tip
by measuring the cutoff in the simulated CMD for (i) the full stellar
population mix (representative of the Fornax RGB), and (ii) only stars
older that 10 Gyr. As a consequence of the adopted metal enrichment
history, the TRGB in the overall synthetic CMD turns out to be
{\it brighter} than that for old stars alone.
The difference in the TRGB $K$ magnitude is 
$\Delta M_K = M_K^{\rm old} - M_K^{\rm all} = 0.22$ mag. 

By taking this population correction into account, the distance modulus
of Fornax was calculated using \cite{vale+2004b} empirical calibration
of the $M_K^{\rm TRGB}$ as a function of metallicity, based on 
Galactic globular clusters:
\begin{equation}
 M_K^{\rm TRGB}=-6.92 - 0.62~{\rm [M/H]} 
\label{e_trgb}
\end{equation} 
where \mh\ is an estimate of the average metal abundance. The
r.m.s. uncertainty of this calibration is $0.16$ mag. We then apply
Eq.~\ref{e_trgb} to the {\it old} stellar population in Fornax:
\begin{equation}
(m-M)_0 = (K^{\rm TRGB} + \Delta M_K)  - A_K - M_K^{TRGB}
\end{equation}
where $K^{\rm TRGB}$ is the measured TRGB level and $A_K$ is the
extinction.  Assuming a low metallicity \mh$=-1.60\pm0.15$ appropriate
for the old population of Fornax \citep{berswood2002,grec+2006mex}, the
corrected distance modulus is
$(m-M)_0=20.75\pm 0.19$, 
where the error is dominated by the uncertainty on the
\cite{vale+2004b} calibration.  

It is interesting that, once the stellar content is properly taken into
account, the $K$-band RGB tip provides an estimate of the distance to
Fornax in good agreement with our determination from the RC and other
estimates in the literature.  However, the uncertainty remains high,
given the strong dependence of the TRGB on age and metallicity and the
error on the calibration relation.

In summary, the distance modulus derived here from near-infrared data
and methods appears to confirm those measured by \citet{savi+2000} from
the $I$ magnitude of the RGB tip, \dmod~$= 20.70$, and from the mean
magnitude of old HB stars, \dmod~$= 20.76$. New distance measurements
based on wide-field optical observations are presented by
\citet{rizz+2006}, where the different results in the literature are
compared and discussed in some detail.

\section{Metallicity}
\label{s_met}

\begin{table} 
\caption{The mean metallicity of Fornax from photometric indices of 
\cite{vale+2004a} and \cite{kuchfrog1995}.}\label{t_fe}
%
\begin{tabular}{l c c }
\hline\hline 
index &\feh$_{\rm GC}$ & \mh\\
\hline 
$(J-K)_0^{M_K=-5.5}$	& $  -0.98\pm 0.18   $ & $  -0.81\pm 0.17 $ \\
$(J-K)_0^{M_K=-5.0}$	& $  -0.97\pm 0.20   $ & $  -0.83\pm 0.19 $ \\
$(J-K)_0^{M_K=-4.0}$	& $  -0.98\pm 0.19   $ & $  -0.86\pm 0.18 $ \\
$(J-K)_0^{M_K=-3.0}$	& $  -1.01\pm 0.15   $ & $  -0.79\pm 0.14 $ \\

$(V-K)_0^{M_K=-5.5}$	& $  -1.13\pm 0.15   $ & $  -0.96\pm 0.16 $ \\
$(V-K)_0^{M_K=-5.0}$	& $  -1.23\pm 0.27   $ & $  -1.02\pm 0.25 $ \\
$(V-K)_0^{M_K=-4.0}$	& $  -1.37\pm 0.39   $ & $  -0.94\pm 0.35 $ \\
$(V-K)_0^{M_K=-3.0}$ 	& $  -1.28\pm 0.38   $ & $  -1.04\pm 0.39 $ \\

VFO slope$_{\rm RGB}$	& $  -1.20  $ & $  -1.03$ \\
KF slope$_{\rm RGB}$    & $  -1.16  $ & \nodata   \\
\hline 
\end{tabular}
\end{table}

\subsection{Mean metallicity}

The mean metallicity of the stellar populations that make up the
Fornax RGB were estimated by comparing the near-infrared and
optical-infrared colors of red giants to the RGB fiducial lines of
Galactic globular clusters of known metal abundance.
\cite{vale+2004a} calibrated the RGB colors at fixed $K$-band
luminosities in the near-infrared CMDs of Milky Way globular clusters as
a function of metallicity.  They give color-metallicity relations for
$(V-K)_0$ and $(J-K)_0$ at $M_K=-5.5, -5.0, -4.0, -3.0$ against the
[Fe/H] of \abbrev{GGCs} on the scale of \citet{carrgratt1997}.

They also provide calibrations against [M/H], a mean metallicity
measuring the abundance of all heavy elements.  This parameter is
particularly important to estimate the metallicities of dwarf spheroidal
galaxies by comparison with the photometric properties of Milky Way
globular clusters. These objects are known to show non-solar abundance
patterns, with an overabundance of $\alpha$-elements relative to iron
that is a function of the cluster metallicity 
\citep[][ and references therein]{prit+2005}. 
In contrast, dwarf spheroidal galaxies tend to have
[$\alpha$/Fe] ratios closer to solar \citep[e.g., ][]{shet+2003}. 
\citet{sala+1993} have shown that the color of red giant stars is driven
by the overall metal abundance rather than the Fe abundance.  Thus, the
{\it iron} [Fe/H] scale of Galactic globular clusters is not immediately
applicable to dwarf galaxies. Instead, the [M/H] provides a suitable
parameter for comparing stellar systems with different abundance
patterns and rank them againsts Milky Way globular clusters.  For the
sake of comparison with previous works, we provide here both the [Fe/H]
and \mh\ rankings, but recommend the \mh\ as the most appropriate
values.

The mean [Fe/H] and $\mh$ values of Fornax RGB stars computed using the
\cite{vale+2004a} calibrations are presented in Table~\ref{t_fe},
together with errors calculated from color uncertainties by error
propagation. The average values are 
$\feh_{V-K} = -1.22 \pm0.12$ and
$\feh_{J-K} = -0.99\pm0.09$ \citep[on the scale of ][]{carrgratt1997}.  Using the
calibrations against \mh, we obtain a mean metallicity 
$\mh_{V-K} = -0.98 \pm0.12$ and
$\mh_{J-K} = -0.82\pm0.08$
Alternatively, a robust metallicity indicator is represented by the RGB
slope.  The \cite{kuchfrog1995} calibration yields
$\feh_\text{ZW}=-1.28$ on the 
\citet[][ \abbrev{ZW84}]{zinn+west1984} scale, corresponding
to $\feh=-1.16$ on the \abbrev{CG97} scale \citep[using the conversion
of ][]{carr+2001}.  This value is confirmed by the recent re-calibration
of the slope by \cite{vale+2004a}, yielding $\feh=-1.20$ or \mh~$=
-1.03$, in good agreement with other methods.

\subsection{Metallicity distribution}

\begin{table} 
\caption{Galactic globular clusters used as RGB metallicity
templates. We list metallicities on the \cite{zinn+west1984} and
\cite{carr+2001} [Fe/H] scales, as well as ``global'' metallicities
$\mh$ (see text).  
}\label{t_gcs}
\begin{tabular}{l c c c}
\hline\hline 
cluster & 
\null~~~$\feh_\text{ZW}$&
\null~~~$\feh_\text{CG}$&
\mh\\
\hline 
M\,15           & $ -2.17 $ & $ -2.12 $ & $ -1.91$ \\
M\,68           & $ -2.09 $ & $ -1.99 $ & $ -1.81$ \\
M\,55           & $ -1.82 $ & $ -1.61 $ & $ -1.41$ \\
M\,13           & $ -1.65 $ & $ -1.39 $ & $ -1.18$ \\
NGC\,288        & $ -1.40 $ & $ -1.07 $ & $ -0.85$ \\
NGC\,362        & $ -1.27 $ & $ -1.15 $ & $ -0.99$ \\
47\,Tuc         & $ -0.71 $ & $ -0.70 $ & $ -0.59$ \\
NGC\,6624       & $ -0.63 $ & $ -0.63 $ & $ -0.48$ \\
NGC\,6553       & $ -0.29 $ & $ -0.06 $ & $ -0.05$ \\
NGC\,6528       & $ -0.33 $ & $ +0.07 $ & $ +0.05$ \\
\hline 
\end{tabular}
\end{table}

The metallicity of RGB stars can also be estimated by interpolating the
colors of individual stars across a grid of empirical RGB templates.  In
this way, we also obtain an observable related to 
the metallicity distribution function
(\abbrev{MDF}) of the red giant stars.
The method used here \citep[see ][]{savi+2000b}
consists in building up a family of hyperbolae that best fit the RGB
fiducial lines of GGCs of known metallicity, and then use interpolation
over that family of curves to derive the metallicity of RGB stars.
A previous application of the method to the $(V-K)$, $M_K$ plane can be
found in \cite{zocc+2003}, \refereeEVH{where it is discussed in some
detail} for stars in the Galactic bulge.

Table~\ref{t_gcs} lists the reference globular clusters with their
adopted metallicities.  In the present implementation of the method we
have employed $\mh$ values from \cite{vale+2004a}, with the updated
$\feh_{\text{CG}}$ scale of \citet{carr+2001}.

The resulting ``photometric MDF'' is shown in Fig.~\ref{f_mdf} along
with an illustration of the interpolation method.  
We also estimated the statistical uncertainty of our photometric
metallicity determinations using a Monte Carlo approach.  A synthetic CMD
was generated by randomly choosing 10\,000 stars along the best fit line
in Fig. \ref{f_mdf}.  Errors were added to the synthetic magnitudes
according to a Gaussian distribution with standard deviations 
0.01 in $V$ and 
\referee{according to the results of artificial star experiments in
$K$ (see Sect. \ref{s_compl}).}
We then used the same
algorithm applied to the Fornax CMD to retrieve individual metallicities.
The recovered metallicities have a nearly Gaussian distribution 
with a dispersion $\sigma=0.06$ dex. This scatter was taken
as the statistical uncertainty associated to our individual
metallicities.  
\refereeEVH{We note that the photometric completeness in the region of
the CMD used to compute the MDF is 100\%, so that no completeness
correction is needed.}

\realfigure{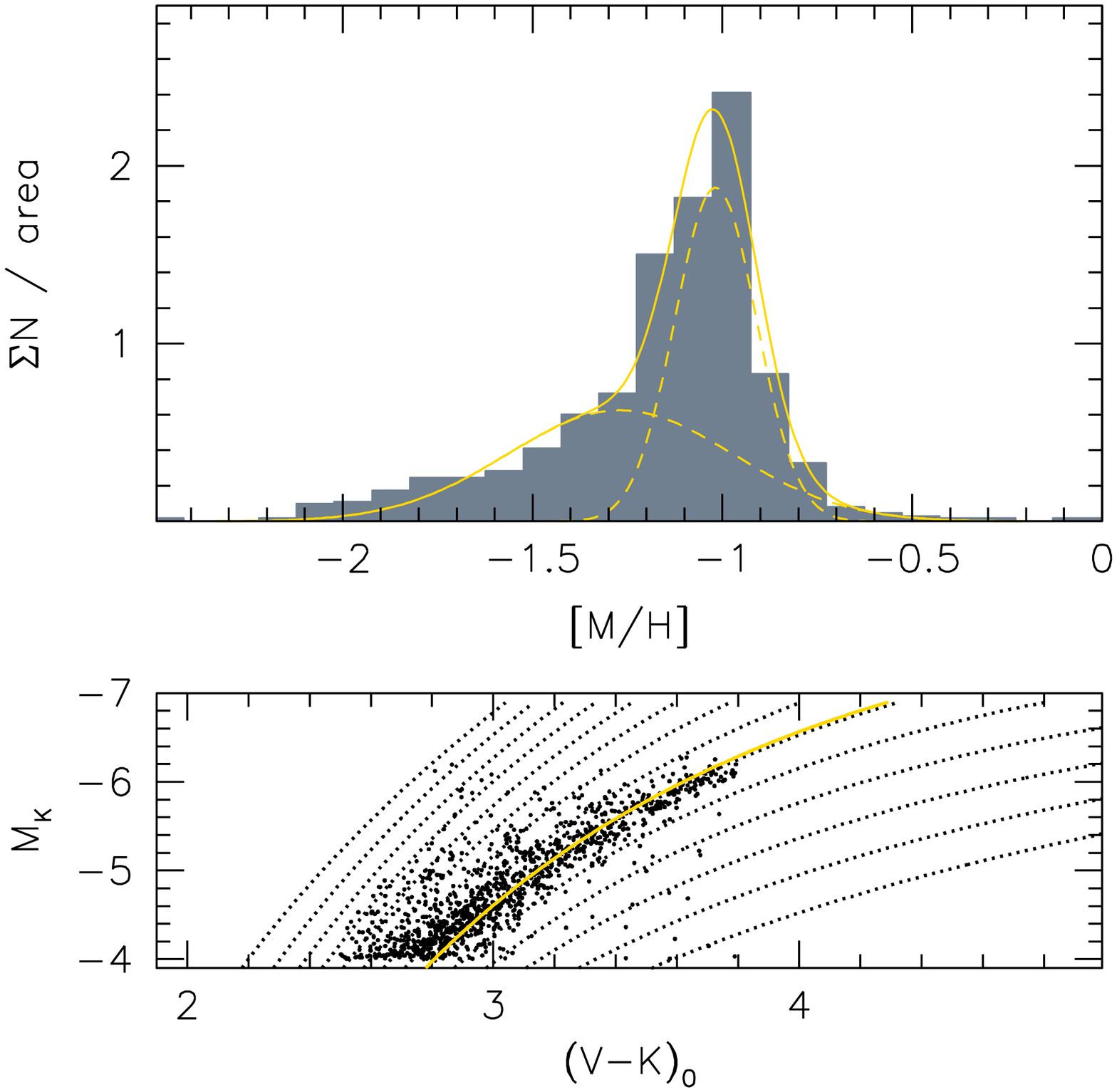}{Derivation of a photometric Metallicity
Distribution Function of RGB stars in Fornax using $(V-K)$ colors.
{\it{Upper panel:}} the photometric metallicity distribution function of
red giants in Fornax. The {\it solid line} is a fit of the sum of two
Gaussian functions ({\it dashed lines}) to the data.
The method is illustrated in the {\it{lower panel}}: the stars on the
upper RGB of Fornax are interpolated across analytical fits ({\it dotted
lines}) to the RGB fiducial lines of template Galactic globular
clusters. The {\it yellow line} marks the best-fit mean metallicity.
}{f_mdf}

The optical--near-IR
data confirm the extended metallicity distribution of Fornax stars
suggested by \citet{savi+2000}.  The distribution is formally modeled by
the sum of two Gaussians, with a main peak at $\mh \simeq -1.02$ and a
secondary peak at $\mh \simeq -1.28$.  The mode corresponds to
about $\feh \simeq -1.2$ and $-1.3$ on the CG97 and ZW84 scales,
respectively.
\referee{The standard deviation of the Gaussian corresponding to the
main peak is $0.11$ dex, which is much larger than the scatter of $0.06$
dex due to photometric error.}
We remind that our ``metallicity distribution'' is representative of the
MDF only for old stellar populations with age comparable to the old
globular clusters in the Milky Way.  This is certainly not true for
Fornax stars, whose mean age is about 5-6 Gyr \citep{savi+2000}.

In all cases, however, this distribution represents an important
observable that models of galactic evolution should be able to
reproduce.

\subsection{Age correction}

\realfigure{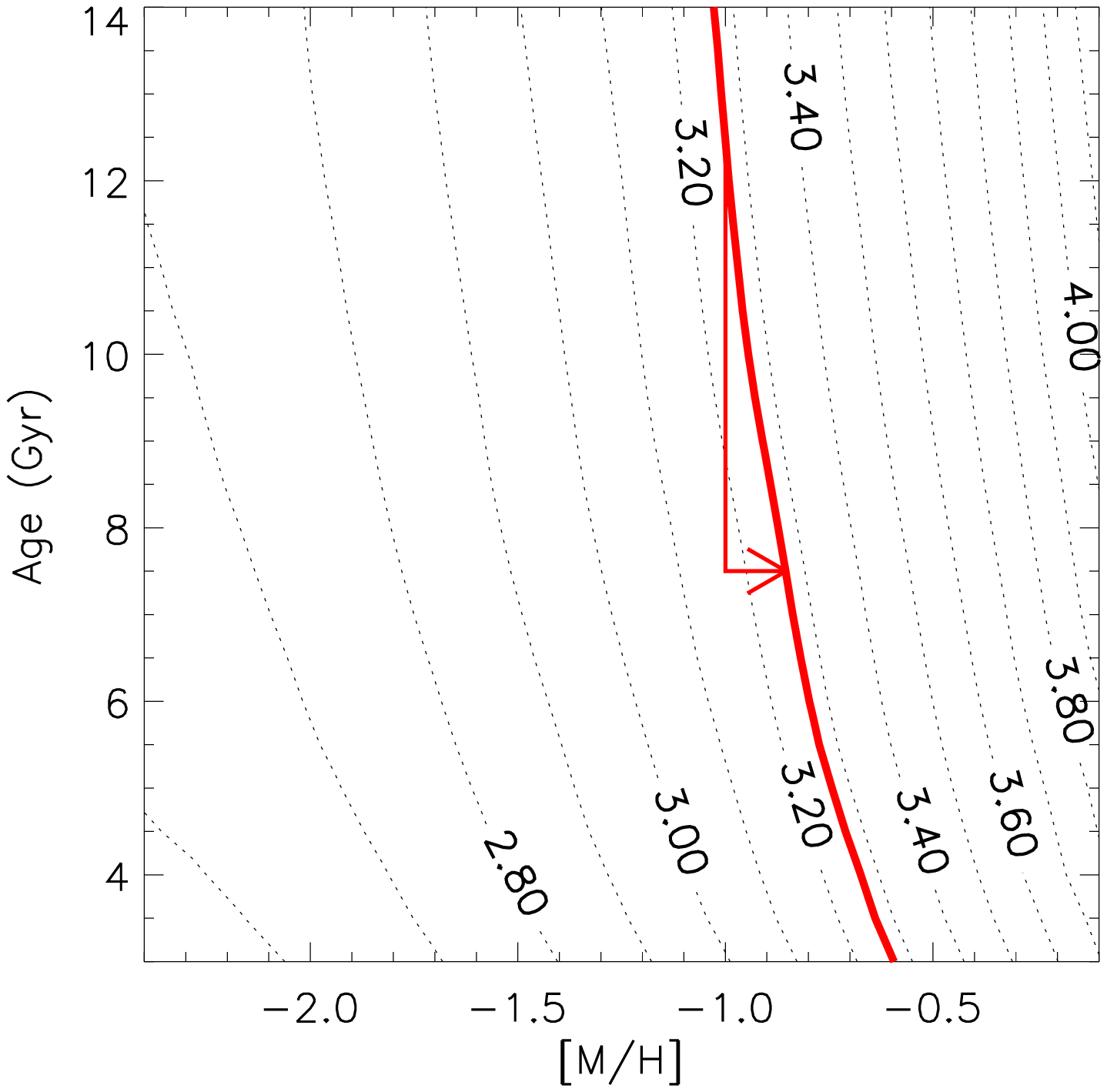}{Contours of constant $(V-K)$ color at
$M_K=-5.0$ as a function of age and [M/H] ({\it dotted lines}), using
the model isochrones of \cite{piet+2004}. This plot allows a
quantification of the age-metallicity degeneracy in the metallicity
estimates based on optical-infrared colors.
The {\it red solid line} corresponds to the color of the RGB of
Fornax. An arrow shows the change in metallicity at this fixed color
between 12.5 Gyr old globular clusters and a 7.5 Gyr old galaxy
(representative of Fornax \abbrev{dSph}).  }{f_agecor}

Since the bulk of the stellar populations in Fornax is younger than
stars in GGCs, the red giants stars in Fornax are on average slightly
bluer than globular cluster stars of the same metallicity.  As a
consequence, the metallicity obtained from the mean RGB color is
systematically underestimated.
Using optical--near-infrared colors, however, this ``age-metallicity
degeneracy'' is much reduced with respect to optical colors. This is
illustrated in Fig.~\ref{f_agecor}, where we show the effects of age and
metallicity variations on RGB colors at $M_K=-5.0$ using the
\cite{piet+2004} isochrones.
In this figure, contour lines of equal color are nearly vertical.  This
means that the effects of a change in age (or those of an age spread) on
the color shift and the color dispersion are {\rm much smaller} than
those produced by metallicity variations.

We have used this plot to estimate a mean correction to metallicity, by
assuming that the age of \abbrev{GGCs} and Fornax dSph are 12.5 and 7.5
Gyr, respectively.  The correction is the difference in metallicity
needed to keep the color constant while moving from 12.5 Gyr to 7.5
Gyrs.
We found a correction $\Delta \mh \simeq 0.15$, yielding for Fornax an
age-corrected metallicity 
$\mh=-0.87$ in terms of global metallicity (we consider the mode of the
photometric MDF).

This result is in good agreement with previous results in the
literature.  \citet{savi+2000} derived a mean age-corrected metallicity
$\feh=-1.0$ from the $(V-I)$ color of the RGB, on the ZW84 scale
(corresponding to [Fe/H]$_{\rm CG} \approx -0.9$).

\subsection{Comparison with spectroscopy}

\realfigure{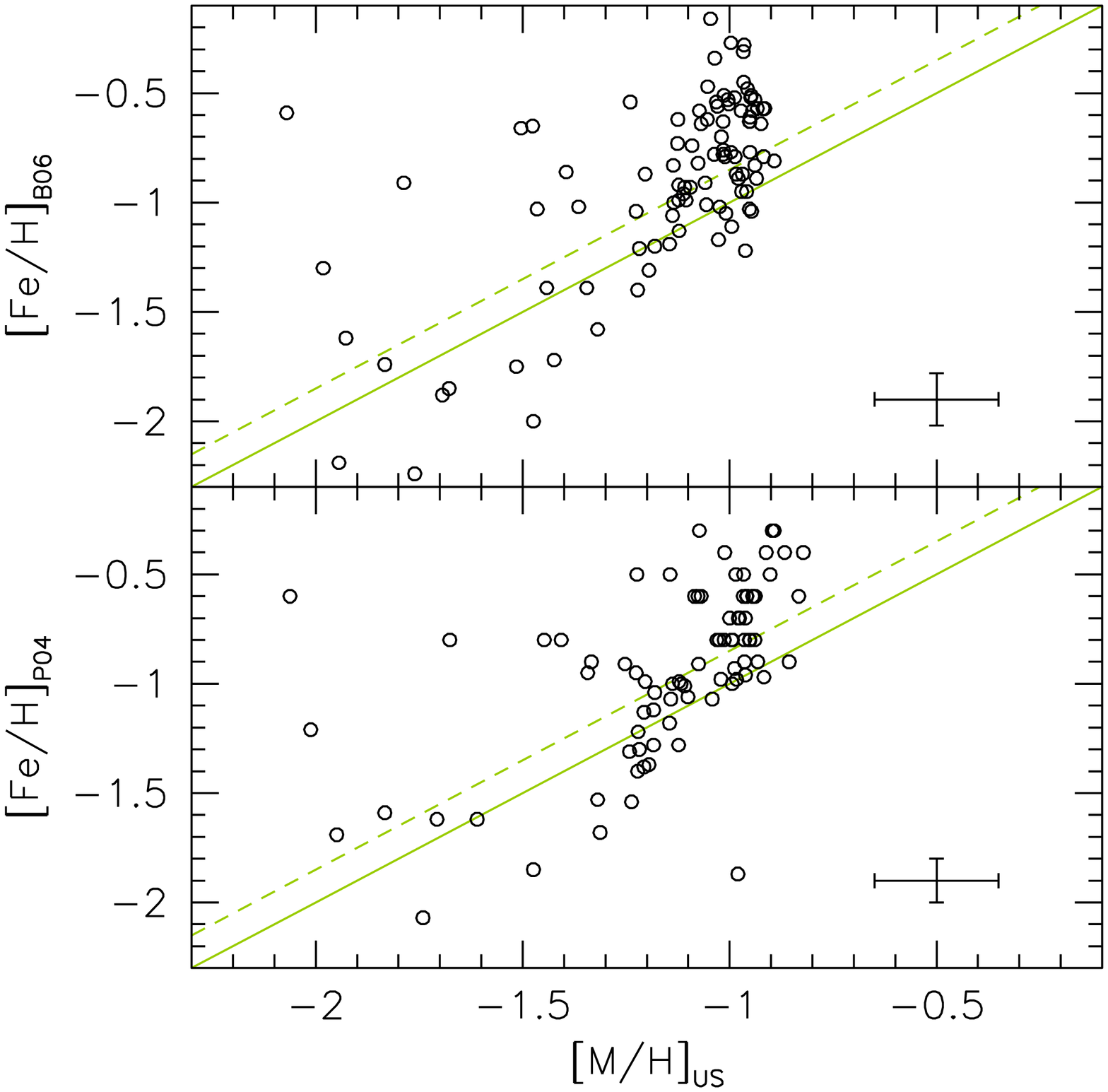}{
Comparison of metallicities of
individual stars inferred from near-infrared colors and metal abundances
derived from the strengths of \caii\ triplet lines.  {\it Lower panel:}
spectroscopic data from \citet{pont+2004}; {\it upper panel:} data from
\citet{batt+2006}. The {\it solid} line indicates one-to-one
correlation, while a {\it dashed} line shows the shift expected by
taking into account the mean age of stellar populations in Fornax.
Error crosses represent the mean uncertainties of spectroscopic measures,
and the $3\sigma$ error of our photometric determination.
}{f_conf_pont}

Our mean metallicity agrees well with the results of low-resolution
spectroscopic analyses of RGB stars, yielding a mean metallicity
$\feh=-1.0$ and $\feh=-0.9$ \citep{tols+2001,pont+2004} on the CG97
scale. This value has been confirmed by high-resolution
spectroscopy of (two) stars by \citet{tols+2003}.
Since the extended star formation of Fornax leads
to an abundance ratio ${\rm [\alpha/Fe]} \simeq 0$
\citep{shet+2003}, we can assume that our \mh\ values are
directly comparable with the spectroscopic measurements.

The relatively large overlap between our optical--near-IR photometry and
the spectroscopic samples of Fornax RGB stars with metallicities derived
by \caii\ triplet spectroscopy \citep{pont+2004,batt+2006} allows us a
direct comparison of metal abundances  on a star-by-star basis.
This comparison is especially interesting to assess the reliability of
photometric \abbrev{MDF} determinations for all systems that are too
distant for (even low-resolution) spectroscopy.  

Figure~\ref{f_conf_pont} plots the metallicities of individual stars
derived from $(V-K)$ colors against those estimated from the
equivalent widths of the \caii\ triplet lines by \citet{pont+2004}
(lower panel) and \citet{batt+2006} (upper panel). 
For the comparison with \citet{pont+2004} results we 
used only spectra with noise below a
given threshold (F. Pont, priv. comm.).  The spectroscopic values for
metal-rich stars are those corrected by the authors by comparison with
high-metallicity stars in the LMC \citep[see][ their section 3]{pont+2004}.
While a few stars appear too red or too blue for their
spectroscopic metallicity, the general trend is that of an overall
correlation. A discrepancy is apparent at the metal-rich end, where the
metallicity estimates from photometry appear to saturate.
A similar trend, with a worse correlation, is noticed in the comparison
with \citet{batt+2006} data, where there is a large excess of metal-rich
stars with respect to the photometric estimates.

\realfigure{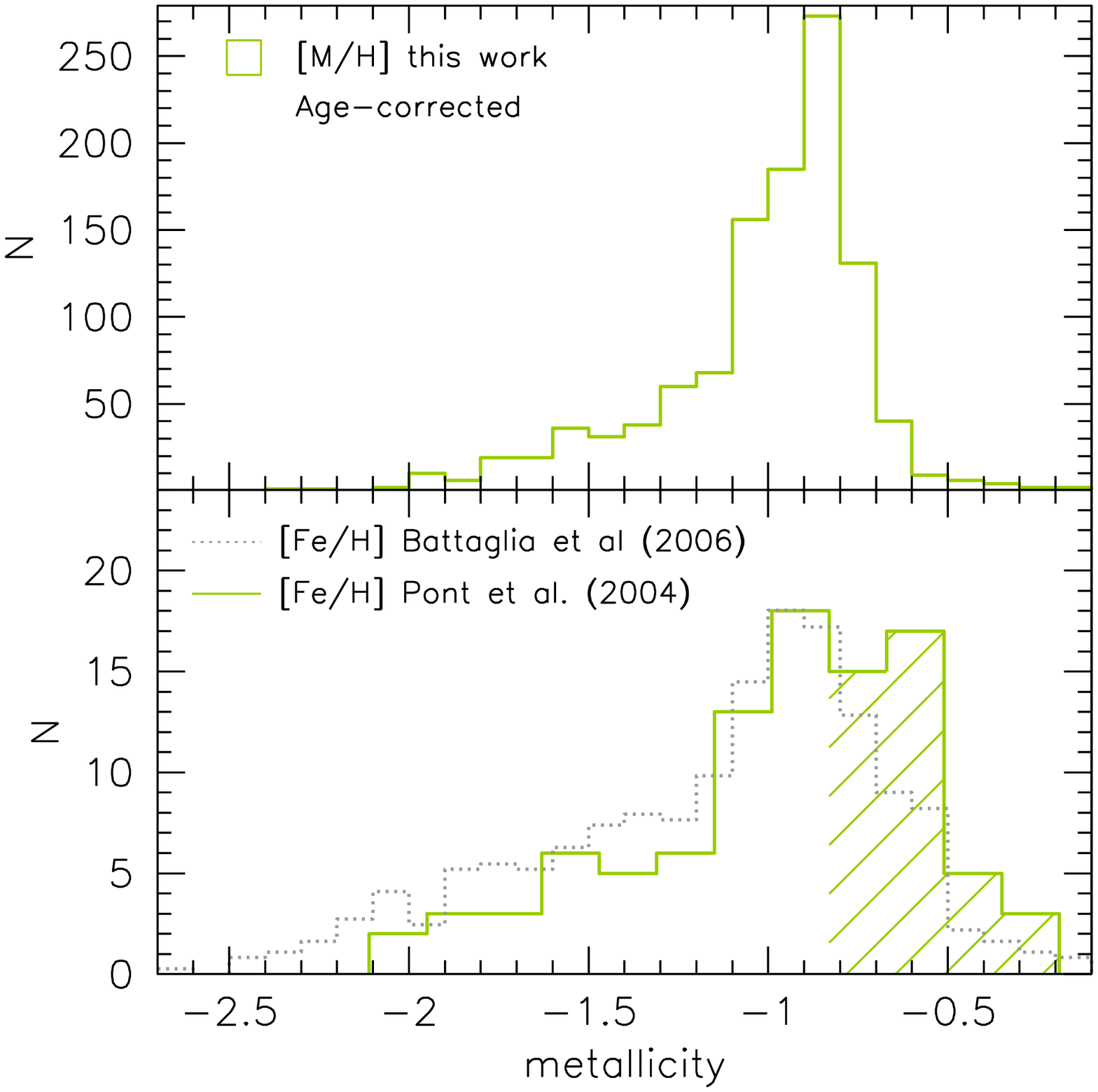}{Comparison of the
metallicity distribution inferred from the optical-infrared colors of
RGB stars (this paper, {\it upper panel}) and the MDF obtained from
\caii\ triplet spectroscopy \citep[][ {\it lower
panel}]{pont+2004,batt+2006}. The shaded part of the histogram from
\citet{pont+2004} ([Fe/H]~$> -0.8$) contains extrapolated metallicities.
A mean age correction, resulting in a shift of 0.15 dex, has been
applied to our histogram.
}{f_2hist}

The metallicity distribution of Fornax RGB stars is shown in
Fig.~\ref{f_2hist}, together with the MDFs obtained from \caii\
spectroscopy. 
In this case, the distribution of photometric [M/H] values in
Fig.~\ref{f_2hist} was corrected by 0.15 dex toward higher metallicities
to take into account the fact that intermediate age stars (the bulk of
Fornax stars) are bluer than the \abbrev{GGC} template stars of the same
metallicity.  Although this correction clearly represents a first order
approximation, it is interesting to note the agreement in the {\it mode}
of the metallicity distribution with the results of spectroscopy.

Clearly, the first explanation for the discrepancy at the metal-rich end
is the age-metallicity degeneracy in RGB star colors that even
optical-infrared color indices cannot completely overcome.
Assuming that the \caii\ triplet methods provides the correct
metallicities, the behavior noticed in Fig.~\ref{f_conf_pont} 
and \ref{f_2hist} appears to
be consistent with a late metal enrichment scenario suggested by the
cited spectroscopic studies.  Indeed, the redder optical-infrared colors
of a young population of metal-rich red giants are compensated for by
the younger age.
However, spectroscopic abundances from \caii\ triplet line strengths are
also somewhat uncertain, especially in the high-metallicity regime
(where an extrapolation may be needed) and when the \caii\ triplet
calibration, which is based on globular cluster stars, is applied to the
spectra of young (1-4 Gyr old)  stars. Accurate
abundance measurements from high-resolution spectroscopy may be useful
to definitively clarify the issue.

\section{Summary and conclusions}
\label{s_sum}

We have presented a near-infrared photometry of the stars in the Fornax
dwarf spheroidal galaxy. Our study provides color-magnitude diagrams and
photometric catalogs of red giant and AGB stars in Fornax over a
$18\farcm5 \times 18\farcm5$ area, and deep photometry over a $4\farcm5
\times 4\farcm5$ central field.
The main results are the following:

\begin{itemize}

\item{From stars on the red giant branch and the \abbrev{RC}, we have
obtained independent estimates of the distance to Fornax based on mean
$K$ magnitude of the red clump and the RGB tip, which take into
account the mean age of the stellar populations in Fornax.  The average
value obtained from the two methods is \dmod$ = $\distfor, in excellent
agreement with previous authors, and in particular with the results of
\citet{savi+2000} from optical photometry.}

\item{The $(V-K)$ color distribution of RGB stars has been used to infer
the mean metallicity and metallicity distribution of red giant stars in
Fornax, taking advantage of the reduced dependence of $(V-K)$ colors of
RGB stars from the age of the stellar population.  The average
metallicity was found to be 
$\mh \simeq -0.9$. This compares well with the values recently
obtained from spectroscopy by Tolstoy et al.  (\citeyear{tols+2001},
\citeyear{tols+2003}), \citet{pont+2004}, and \citet{batt+2006}.}

\item{The metallicity distribution is consistent with that obtained
from spectroscopy up to the metallicity of 47\,Tuc ($\feh \sim -0.7$).
However, there is a clear discrepancy between the \abbrev{MDFs} derived
from near-infrared colors and spectroscopy near the metal-rich end,
where \citet{pont+2004} found of a tail of stars with metallicity up to
almost solar.
This discrepancy could be caused by the effects of the age-metallicity
degeneracy, which cannot be completely corrected even using 
near-IR photometry, but
the extrapolation used by \citet{pont+2004} to derive their
metallicities from the measurements of \caii\ lines, as noted by the
authors could also have some effect.  More observations are needed to
solve this discrepancy and establish the upper end of metal enrichment
in Fornax.}

\end{itemize}

\begin{acknowledgements}
We are indebted with F. Pont, G. Battaglia and collaborators for kindly
providing unpublished information about their spectroscopic results. We 
thank M. Salaris for helpful discussions of the properties red
clump stars.
We also thank an anonymous referee for comments and suggestions 
that improved the presentation of the paper.
M.G. and E.V.H. acknowledge support by MIUR, under the scientific
projects PRIN\,2002028935 and PRIN\,2003029437.
This publication makes use of data products from the Two Micron All Sky
Survey, which is a joint project of the University of Massachusetts and
the Infrared Processing and Analysis Center/California Institute of
Technology, funded by the National Aeronautics and Space Administration
and the National Science Foundation.
\end{acknowledgements}

\bibliography{bib_irfornax}
\bibliographystyle{aa}

\end{document}